\pdfoutput=1
\newcommand*{\ATLASLATEXPATH}{latex/}
\documentclass[cernpreprint,UKenglish,texlive=2016,LANGEDIT=true]{\ATLASLATEXPATH atlasdoc}

\DeclareOldFontCommand{\rm}{\normalfont\rmfamily}{\mathrm}
\DeclareOldFontCommand{\sf}{\normalfont\sffamily}{\mathsf}
\DeclareOldFontCommand{\tt}{\normalfont\ttfamily}{\mathtt}
\DeclareOldFontCommand{\bf}{\normalfont\bfseries}{\mathbf}
\DeclareOldFontCommand{\it}{\normalfont\itshape}{\mathit}
\DeclareOldFontCommand{\sl}{\normalfont\slshape}{\@nomath\sl}
\DeclareOldFontCommand{\sc}{\normalfont\scshape}{\@nomath\sc}

\usepackage[]{\ATLASLATEXPATH atlaspackage}
\usepackage{\ATLASLATEXPATH atlasbiblatex}

\usepackage{\ATLASLATEXPATH atlascontribute}

\usepackage{\ATLASLATEXPATH atlasphysics}

\usepackage{\ATLASLATEXPATH atlascover}

\graphicspath{{logos/}{figures/}}

\usepackage{EXOT-2017-18-PAPER-defs}

\AtlasTitle{
    \Large{Search for heavy Majorana or Dirac neutrinos and
    right-handed $W$ gauge bosons in final states with two charged leptons
    and two jets at $\sqrt{s} = \SI{13}{\TeV}$ with the ATLAS detector}
}
\author{The ATLAS Collaboration}

\AtlasRefCode{EXOT-2017-18}

\PreprintIdNumber{CERN-EP-2018-199}

\AtlasJournal{JHEP}
\AtlasJournalRef{J. High Energ. Phys. (2019) 2019: 16}
\AtlasDOI{10.1007/JHEP01(2019)016}

\AtlasAbstract{%
A search for heavy right-handed Majorana or Dirac neutrinos \NR and heavy right-handed gauge bosons \WR is
performed in events with a pair of energetic electrons or muons, with the same or opposite electric charge, and two energetic jets. 
The events are selected from $pp$ collision data with an integrated luminosity of \SI{36.1}{\ifb} collected by the
ATLAS detector at $\sqrt{s} = \SI{13}{\TeV}$. No significant deviations from the Standard Model are observed.
The results are interpreted within the theoretical framework of a left--right
symmetric model and lower limits are set on masses in the heavy right-handed $W$ boson and neutrino mass plane.
The excluded region extends to \mWR~$= \SI{4.7}{\TeV}$ for both Majorana and Dirac \NR neutrinos.
}

\hypersetup{pdftitle={
    Search for heavy Majorana or Dirac neutrinos and
    right-handed $W$ gauge bosons in final states with two charged leptons
    and two jets at $\sqrt{s} = 13~\TeV{}$ with the ATLAS detector
    },
    pdfauthor={The ATLAS Collaboration}
}

\begin{document}

\LEcontact{Pippa Wells pippa.wells@cern.ch}

\maketitle

\tableofcontents
\newpage


\section{Introduction}
\label{sec:intro}
Left--right symmetric models \cite{Pati:1974yy, Mohapatra:1974hk, Mohapatra:1974gc, Senjanovic:1975rk, Senjanovic:1978ev, Minkowski:1977sc, Mohapatra:1979ia, Marshak:1979fm, Mohapatra:1980qe} (LRSMs) attempt to explain the broken parity symmetry of the weak interaction in the Standard Model (SM) and can introduce, depending on the form of the LRSM, right-handed counterparts to the $W$ and $Z$ bosons (\WR and $Z_R$), and right-handed heavy neutrinos (\NR).
A search for \WR boson and \NR neutrino production in a final state
containing two charged leptons and two jets (\lljj) with $\ell = e, \mu$ is presented here.
The exact process of interest is the Keung--Senjanovi\'c (KS) process \cite{Keung:1983uu}, shown in Figure~\ref{fig:KeungSenjanovic}.
When the \WR boson is heavier than the \NR neutrino (\mWR > \mNR), the on-shell \WR mass can be reconstructed from the invariant mass of the \lljj system, whereas, when \mNR > \mWR, the on-shell \WR mass can be reconstructed from the invariant mass of the \jj system.
Only \EE and \MM lepton pairs, coupling respectively to $N_{\text{R}}^e$ and $N_{\text{R}}^{\mu}$,
are considered as part of the \lljj final state, since no mixing between flavours is assumed.  
Left- and right-handed weak gauge couplings are also defined to be equivalent ($g_{\text{L}} = g_{\text{R}}$).

\begin{figure}[h]
\centering
\subfloat[]{
\resizebox{0.4\textwidth}{!}{
  \begin{tikzpicture}[line width=1.25 pt, scale=1]

  \draw[antifermion] (0,1) -- (1,0) ;
  \node at (-0.2,1) {$\bar{q}$};
  \draw[fermion] (0,-1) -- (1,0) ;
  \node at (-0.2,-1) {$q$};

  \draw[vector] (1,0) -- (2,0);
  \node at (1.5,-0.35) {\WR};

  \draw[fermionnoarrow] (2,0) -- (3,1);
  \node at (2.75,0.3) {\NR};
  \draw[fermionnoarrow] (2,0) -- (3,-1);
  \node at (3.2,-1) {$\ell$};

  \draw[vector] (3,1) -- (4,2);
  \node at (3.85,1.25) {\WRstar};
  \draw[fermionnoarrow] (3,1) -- (4,0);
  \node at (4.2,0) {$\ell$};

  \draw[antifermion] (4,2) -- (5,3);
  \node at (5.2,3) {$\bar{q}$};
  \draw[fermion] (4,2) -- (5,1);
  \node at (5.2,1) {$q$};

  \end{tikzpicture}
}
\label{fig:KeungSenjanovic1}
}
\subfloat[]{
\resizebox{0.4\textwidth}{!}{
  \begin{tikzpicture}[line width=1.25 pt, scale=1]

  \draw[antifermion] (0,1) -- (1,0) ;
  \node at (-0.2,1) {$\bar{q}$};
  \draw[fermion] (0,-1) -- (1,0) ;
  \node at (-0.2,-1) {$q$};

  \draw[vector] (1,0) -- (2,0);
  \node at (1.5,-0.35) {\WRstar};

  \draw[fermionnoarrow] (2,0) -- (3,1);
  \node at (2.75,0.3) {\NR};
  \draw[fermionnoarrow] (2,0) -- (3,-1);
  \node at (3.2,-1) {$\ell$};

  \draw[vector] (3,1) -- (4,2);
  \node at (3.85,1.25) {\WR};
  \draw[fermionnoarrow] (3,1) -- (4,0);
  \node at (4.2,0) {$\ell$};

  \draw[antifermion] (4,2) -- (5,3);
  \node at (5.2,3) {$\bar{q}$};
  \draw[fermion] (4,2) -- (5,1);
  \node at (5.2,1) {$q$};

  \end{tikzpicture}
}
\label{fig:KeungSenjanovic2}
}
\caption{The KS process, for \protect\subref{fig:KeungSenjanovic1} the \mWR > \mNR case and \protect\subref{fig:KeungSenjanovic2} the $\mNR~>~\mWR$ case.}
	\label{fig:KeungSenjanovic}
\end{figure}

In the minimal LRSM containing the type-I seesaw mechanism \cite{Minkowski:1977sc, Mohapatra:1979ia, Marshak:1979fm, Mohapatra:1980qe}, \NR neutrinos are Majorana particles.
The type-I seesaw mechanism accounts for the masses of the SM neutrinos by linking (heavy) \NR neutrinos and the SM neutrino
masses through a mixing matrix.
In this case, both the
SM neutrinos and the hypothetical \NR neutrinos are required to be
Majorana particles, allowing lepton-number-violating processes, such as the KS
process, to occur.
In LRSM variants, including the inverse seesaw
mechanism~\cite{Mohapatra:1986aw, Mohapatra:1986bd, Chen:2011hc, Dev:2015pga}, \NR neutrinos are pseudo-Dirac
particles\footnote{A pseudo-Dirac particle is formed by two Majorana particles with identical masses \cite{Wolfenstein:1981kw}.} (referred to in this paper as ``Dirac''
particles for simplicity).  
For simple versions of LRSMs containing the inverse seesaw mechanism, lepton-number-violating processes are not expected \cite{Das:2017hmg}.
The Majorana or Dirac nature of the \NR neutrino can be established by comparing the charges of the two final-state leptons.
If the \NR neutrinos are Dirac particles, the leptons will always have opposite-sign (\OC) charges.
However, if they are Majorana particles, the \NR neutrinos are their own
anti-particles, and their decay will give rise to an
experimental signature of both
opposite- and same-sign (\SC) dileptons in a 50\%-50\% admixture.

The KS process resulting in an \lljj final state in the
electron and muon channels has been studied
by both the ATLAS and CMS collaborations using $\sqrt{s}=\SI{7}{\TeV}$ \cite{EXOT-2011-14, CMS-EXO-11-091}, $\sqrt{s}=\SI{8}{\TeV}$ \cite{EXOT-2012-24, CMS-EXO-13-008} and $\sqrt{s}=\SI{13}{\TeV}$ \cite{CMS-EXO-17-011} collision data.
CMS also has results for the hadronic $\tau\tau jj$ final state at
$\sqrt{s}=\SI{13}{\TeV}$ \cite{CMS-EXO-16-016, CMS-EXO-16-023}.
No evidence of a \WR boson or a \NR neutrino has emerged from these studies.
The current most stringent exclusion limits on \WR boson and Majorana \NR neutrino
masses are derived by the CMS experiment~\cite{CMS-EXO-17-011} at $\sqrt{s}=\SI{13}{\TeV}$,
in both the electron and muon channels, regions extending to \mWR~$\sim$\SI{4.4}{\TeV}
(for a range of \mNR values) are excluded, whilst the \mNR limits reach $\sim$\SI{2.9}{\TeV} in the electron channel (for \mWR~$\sim$\SI{3.8}{\TeV}) and $\sim$\SI{3}{\TeV} in the muon channel (for \mWR~$\sim$\SI{3.6}{\TeV}).
The analysis presented here extends the ATLAS searches to \SI{36.1}{\ifb} of data at $\sqrt{s}=$\SI{13}{\TeV} and also 
provides lower limits on the masses for the Dirac case of the \NR
neutrino. The cross-section of the KS process varies, depending on the
masses of the \WR boson and the \NR neutrino, from a few pb to about $3
\times 10^{-4}$ pb for high-mass signal points at the end of the
sensitivity of this analysis.
The analysis uses a new \textsc{MadGraph} generator model
~\cite{alwall2011madgraph,Alwall:2014hca} which overcomes known limitations of 
\textsc{Pythia}~\cite{Sjostrand:2007gs} as employed in previous
searches by ATLAS and CMS, and also allows the exploration of the scenario in
which the \NR neutrino is heavier than the \WR boson for the first time.

\FloatBarrier

\section{ATLAS detector}
\label{sec:atlas}
\newcommand{\AtlasCoordFootnote}{
ATLAS uses a right-handed coordinate system with its origin at the nominal interaction point (IP)
in the centre of the detector and the $z$-axis along the beam pipe.
The $x$-axis points from the IP to the centre of the LHC ring,
and the $y$-axis points upwards.
Cylindrical coordinates $(r,\phi)$ are used in the transverse plane,
$\phi$ being the azimuthal angle around the $z$-axis.
The pseudorapidity is defined in terms of the polar angle $\theta$ as $\eta = -\ln \tan(\theta/2)$.
The angular distance is measured in units of $\Delta R \equiv \sqrt{(\Delta\eta)^{2} + (\Delta\phi)^{2}}$.
}

The ATLAS experiment~\cite{PERF-2007-01} is a multipurpose particle detector
with a forward--backward symmetric cylindrical geometry and a near $4\pi$ coverage in
solid angle\footnote{\AtlasCoordFootnote}.
It consists of an inner tracking detector (ID) surrounded by a thin superconducting solenoid
providing a \SI{2}{\tesla} axial magnetic field, electromagnetic (EM) and hadron calorimeters, and a muon spectrometer (MS).
The ID covers the pseudorapidity range $|\eta| < 2.5$.
It consists of silicon pixel, silicon microstrip, and straw-tube transition-radiation tracking detectors.
A new innermost layer of pixel detectors~\cite{ATLAS-TDR-19,Abbott:2018ikt} was installed prior to the start of data taking in 2015.
Lead/liquid-argon (LAr) sampling calorimeters provide electromagnetic energy measurements
with high granularity and cover $|\eta| < 3.2$.
A hadron (steel/scintillator-tile) calorimeter covers the central pseudorapidity range ($|\eta| < 1.7$).
The endcap and forward regions are instrumented with LAr calorimeters
for EM and hadronic energy measurements up to $|\eta| = 4.9$.
The MS surrounds the calorimeters and is based on
three large air-core toroidal superconducting magnets with eight coils each.
The field integral of the toroidal magnets ranges between \num{2.0} and 
6.0 ${\rm T\thinspace m}$
across most of the detector.
The MS includes a system of precision tracking chambers covering $|\eta| < 2.7$ and fast detectors ($|\eta| < 2.4$) for triggering.
A two-level trigger system~\cite{TRIG-2016-01} is used to select events for offline physics analyses. The first-level trigger is
implemented in hardware and uses a subset of the detector information. This is followed by the software-based high-level trigger, reducing the event rate to about \SI{1}{\kHz}.

\FloatBarrier

\section{Dataset and simulated event samples}
\label{sec:simu}
The data used in this analysis were collected at a centre-of-mass energy of \SI{13}{\TeV}
during 2015 and 2016, and correspond to an integrated luminosity of \SI{36.1}{\ifb}\@.
Only high-quality data with all detectors in normal operating conditions are analysed.

Simulated signal and background events are used to optimise the event selection
and to estimate the background contributions.

Signal events with matrix elements calculated by \MGaMC~v2.2.2~\cite{alwall2011madgraph,Alwall:2014hca} were simulated by \PythiaEight~\cite{Sjostrand:2007gs} using the NNPDF2.3~\cite{Ball:2012cx} parton distribution function (PDF) set and the A14 set of tuned parton shower parameters~\cite{ATL-PHYS-PUB-2014-021}.
A version of an LRSM model produced with {\tt FeynRules}~\cite{Alloul:2013bka}
was implemented in \MGaMC \cite{Roitgrund:2014zka} and further modified as described in Ref.~\cite{Nemevsek:2018bbt}.
Events were generated containing only Majorana \NR neutrinos.
For the Dirac neutrino case, only the \OC events are used in the analysis
as no other differences are expected.
Signal samples were generated for different \WR boson and \NR neutrino mass hypotheses,
covering a range from 600~\GeV{} to 5800~\GeV{} for \mWR and 50~\GeV{} to 8000~\GeV{} for \mNR.
Only samples with $\mNR \leq 2\mWR$ are used
as the cross-section for the KS process drops off rapidly with increasing \NR neutrino mass.

For the \OC channel, the dominant SM backgrounds are \Zjet and \ttbar processes.
In the \SC channel, the main backgrounds arise
from misidentified leptons, electron charge misidentification, and 
leptons from diboson processes such as $ZW$, $ZZ$, or $WW$ (including $W^{\pm}W^{\pm}jj$) production.
Other backgrounds taken into the account are $t\bar{t}V$ ($V=W,Z,H$) and single-top production.
Additional rare backgrounds such as the production of four or three top quarks as well as $t\bar{t}WW$,
$t\bar{t}WZ$, $tZ$, and $tWZ$ production, were found to be negligible in both the \OC and \SC channels.

Various SM backgrounds were simulated using different generators and
the dominant backgrounds are evaluated using data-driven techniques, as described
in Section~\ref{sec:background}.

The \Zjet process (with $ Z/\gamma* \to ee/\mu\mu/\tau\tau$) was modelled using \SHERPA 2.2.1~\cite{Gleisberg:2008ta}
with the NNPDF3.0~\cite{ball2015parton} PDF set.
The matrix element (ME) was calculated for up to two partons with next-to-leading-order
(NLO) accuracy in QCD and up to four partons with leading-order (LO) accuracy
using Comix~\cite{Gleisberg:2008fv} and OpenLoops~\cite{Cascioli:2011va}.

For the generation of \ttbar events, \POWHEGBOX v2~\cite{alioli2010general,powheg1,powheg2} was used with the NNPDF3.0 PDF set
in the ME calculation. The parton shower was modelled with
\PythiaEight~\cite{Sjostrand:2007gs} with the NNPDF2.3~\cite{Ball:2012cx} PDF set
and the A14 set of tuned parameters~\cite{ATL-PHYS-PUB-2014-021}.
Single-top-quark events produced in $Wt$ final states were generated with \POWHEGBOX v2
with the CT10 PDF set~\cite{Lai:2010vv} used in the ME calculations.
Single top-quark production via $s$- or $t$-channels was
generated by \POWHEGBOX v1~\cite{alioli2010general,powheg1,powheg2}. This generator uses the four-flavour scheme
for the NLO QCD matrix element calculations together with the fixed four-flavour PDF set CT10f4~\cite{Lai:2010vv}.
The parton shower, fragmentation, and underlying event were simulated with \PythiaSix~\cite{sjostrand2006pythia}
using the CTEQ6L1 PDF sets~\cite{Pumplin:2002vw} and the Perugia 2012 set of tuned parameters~\cite{Skands:2010ak}.
The top-quark mass was set to \SI{172.5}{\GeV}. The EvtGen v1.2.0~\cite{Lange:2001uf} program was used to model bottom
and charm hadron decays.
The $t\bar{t}V$ ($V=W,Z$) processes were generated at LO with \MGaMC~v2.2.2~\cite{alwall2011madgraph}
and the $t\bar{t}H$ process was generated at LO with \MGaMC~v2.3.2. For both, the NNPDF2.3 PDF set was used 
and the parton shower was modelled using \PythiaEight, configured with the A14 tune.

Diboson processes with four charged leptons (4$\ell$), three charged leptons and one neutrino (3$\ell$+1$\nu$),
or two charged leptons and two neutrinos (2$\ell$+2$\nu$) in the final state were generated using \SHERPA 2.2.2
with the NNPDF3.0 PDF set.
The matrix elements considered contained all diagrams with four electroweak vertices.
The various processes were evaluated for up to three partons at LO accuracy and up to one (4$\ell$, 2$\ell$+2$\nu$)
or zero partons (3$\ell$+1$\nu$) at NLO in QCD using the Comix and OpenLoops matrix element generators.
Diboson processes with one boson decaying hadronically and the other one decaying leptonically
were generated with \SHERPA 2.2.1 using the NNPDF3.0 PDF set. The various processes were calculated
for up to three additional partons at
LO accuracy and up to one ($ZZ$) or zero ($WW$, $WZ$) additional partons at
NLO using the Comix and OpenLoops matrix element generators.
Loop-induced and electroweak processes with two weak gauge bosons (W/Z) plus two jets were
simulated with \SHERPA 2.1.1. The calculation is made at LO accuracy in QCD while up to one additional
parton is merged with the matrix element. The CT10 PDF set was used in conjunction with a dedicated parton shower tuning.
The \SHERPA 2.1.1 diboson prediction was
scaled by a factor of $0.91$ to account for differences between the internal electroweak scheme used in this \SHERPA
version and the default scheme.

A single event recorded by the ATLAS detector consists of the superposition of a hard-scattering $pp$ collision
and several additional $pp$ interaction vertices referred to as `pile-up.'
The effect of the pile-up was included by overlaying minimum-bias collisions,
simulated by \PYTHIA~8.186 with a set of tuned parameters referred to as the A2 tune~\cite{ATL-PHYS-PUB-2012-003}
and the MSTW2008LO PDF~\cite{Martin:2009iq}, on each generated signal and background event.
The average number of interactions per $pp$ bunch crossing is 24 and
simulated samples are weighted to reproduce the distribution observed in data.

The detector response for background events was simulated within a framework~\cite{SOFT-2010-01} based on GEANT4~\cite{Agostinelli:2002hh}.
Monte Carlo signal samples were instead processed with a fast simulation~\cite{SOFT-2010-01} which relies on a parameterisation of the calorimeter response~\cite{ATL-PHYS-PUB-2010-013}.
Furthermore, simulated events were processed with the same
reconstruction software used for data. In order to account for the different
particle reconstruction efficiencies measured in data and simulation,
correction factors are derived in dedicated measurements and applied to simulated events.

\FloatBarrier

\section{Object reconstruction}
\label{sec:recon}
The electron and muon selection criteria are summarised in Table~\ref{tab:LepReco}.

\begin{table}[htb!]
\caption{Object reconstruction selection in the \OC and the \SC channels. The requirements corresponding to the
identification and isolation working points are described in Ref.~\cite{ATLAS-CONF-2016-024} under the same names.} 
\label{tab:LepReco} 
\begin{center}
\begin{adjustbox}{width=1\textwidth}
\small
\begin{tabular}{c|c|c|c|c}
\toprule
\footnotesize
Lepton flavour   & \multicolumn{2}{c|}{Electrons} & \multicolumn{2}{c}{Muons} \\
\hline
 Channel             &  OS & SS  & OS & SS \\
\hline
Identification & \multicolumn{2}{c|}{\texttt{LHMedium}}  &  \multicolumn{2}{c}{\texttt{Medium}} \\
\hline
Isolation      & \texttt{LooseTrackOnly} & \texttt{Loose} & \texttt{LooseTrackOnly} &  \texttt{FixedCutTightTrackOnly}\\
\hline
$\pt(\mu)$, $\ET(e)$        & $\ET > 25~\GeV{}$ & $\ET > 30~\GeV{}$ & $\pt > 25~\GeV{}$ & $\pt > 30~\GeV{}$ \\
\hline
$\eta$         & \multicolumn{2}{c|}{$|\eta| < $2.47 and veto $1.37 < |\eta| < 1.52$} & \multicolumn{2}{c}{$|\eta| <$ 2.5}\\
\hline
$|d_0| / \sigma _{d_0}$  & \multicolumn{2}{c|}{ $|d_0| / \sigma _{d_0}< 5$} &  \multicolumn{2}{c}{ $|d_0| / \sigma _{d_0}< 3$} \\
\hline
$|z_0\sin \theta |$  & \multicolumn{4}{c}{$|z_0\sin \theta | < $0.5 mm} \\
\bottomrule
\end{tabular} 
\end{adjustbox}
\end{center} 
\end{table}

Electron candidates are reconstructed from energy deposits in the EM calorimeter which are matched to a track reconstructed
in the ID~\cite{ATL-PHYS-PUB-2016-015}. Candidate electrons must have transverse energy $\ET > 25~(30)~\GeV{}$ in the \OC (\SC) channel, $|\eta| < 2.47$
and satisfy the \texttt{LHMedium} identification criterion based on a multivariate likelihood
discriminant~\cite{ATLAS-CONF-2016-024,ATL-PHYS-PUB-2015-041}. Electrons falling in the transition
region between the barrel and endcap EM calorimeters ($1.37 < |\eta| < 1.52$) are excluded.
The transverse impact parameter ($d_0$) of the track associated with the electron must have a significance 
$ |d_0|/\sigma (d_0) < 5$ relative to the beam line~\cite{STDM-2015-02}, where $\sigma (d_0)$ is the uncertainty
on $d_0$.
All tracks with transverse momentum $\pT > 400~\MeV{}$ associated with electron candidates must have a longitudinal impact parameter multiplied by the sine of the polar angle $|z_0 \sin \theta|$ of less than \SI{0.5}{\mm}.
Electrons are further required to satisfy the track-based (track- and calorimeter-based) isolation criteria
for the \OC (\SC) channel. The isolation selection efficiencies are found to exceed
99\% using $Z \to e^+e^-$ events~\cite{ATLAS-CONF-2016-024}.

Muons are reconstructed from a combined fit of a track in the ID matched to a track in the MS.
The \texttt{Medium} quality criterion, described in Ref.~\cite{PERF-2015-10}, is applied.
Candidate muons must have $\pT > 25~(30)~\GeV{}$ in the \OC (\SC) channel, $|\eta| < 2.5$,
 $|d_0|/\sigma (d_0)<\,\,$3, and $|z_0 \sin \theta|<\SI{0.5}{\mm}$. Muon isolation in the \OC channel is based on track isolation
and has a 99\% efficiency, independent of the muon \pt and $|\eta|$.
In the \SC channel, the track isolation provides an efficiency exceeding 99\% for high-\pt muons
and decreasing to $\sim$95\% for muons with \pt < \SI{40}{\GeV}~\cite{PERF-2015-10}.

The tighter selection criteria applied to leptons used in the SS channel
are designed to suppress the contribution from background events with
misidentified leptons (as discussed in Section~\ref{sec:background}), which are
negligible in the OS channel.

Jets are reconstructed with the \antikt algorithm~\cite{Cacciari:2008gp} using a radius parameter of 0.4 
from energy deposits in the topological clusters of the calorimeter, calibrated
as described in Ref.~\cite{PERF-2016-04}.
Jets must satisfy $\pt > 20~\GeV{}$ and $|\eta|<2.5$. The majority
of jets from pile-up are rejected using the jet-vertex-tagger~\cite{PERF-2014-03}, a
likelihood discriminant combining information from several track-based variables, which is only applied
to jets with $\pt< 60~\GeV{}$ and $|\eta|<2.4$. The tagging efficiency of
pile-up jets with the jet-vertex-tagger is determined
in simulated $Z\rightarrow \mu \mu$ + jet(s) events to be 95\%.

The identification of jets
containing $b$-hadrons (\bjets) is based on a multivariate tagging algorithm~\cite{ATL-PHYS-PUB-2016-012}.
This algorithm uses a set of tracks with loose impact parameter constraints in a
region of interest around each jet axis to enable the reconstruction of the $b$-hadron decay vertex.
The $b$-tagging working point used in the \OC (\SC) channel results in an
efficiency of 70\% (77\%) for jets containing $b$-hadrons. The
expected rejection factors against light-quark and gluon jets, $c$-quark jets and hadronically decaying tau
leptons are respectively 381, 12, 55 for a 70\% efficiency and 134, 6, 22
for a 77\% efficiency. These efficiencies
and rejection factors are determined in a sample of simulated $t\bar{t}$ events~\cite{PERF-2016-05-custom}.

After the electron, muon and jet reconstruction and selection,
possible overlaps between reconstructed objects are resolved.
First, electrons are removed if they share a track with a muon. Next,
ambiguities between electrons and jets are resolved by removing jets
with an angular distance $\Delta R < 0.2$ from electrons;
if $0.2< \Delta R < 0.4$, the electron is removed.
Finally, if a muon and a jet have $\Delta R < 0.4$, the jet is removed 
if it has less than three associated tracks, otherwise the muon is discarded.

\FloatBarrier

\section{Event selection}
\label{sec:evtSelection}
Events were collected using single-lepton or dilepton triggers
during the 2015 and 2016 data-taking periods.
For data collected in 2015, the lowest \ET and lowest \pt trigger thresholds are 24~\GeV{} and 20~\GeV{}
for single-electron and single-muon triggers, respectively.
For 2016 data, the thresholds are 26~\GeV{} and 24~\GeV{} for electron and muon triggers, respectively.
Dielectron triggers with a \ET threshold of 17~\GeV{} are used in the \EESC
case since a different background strategy is considered for the \SC channel (see Section~\ref{sec:background}).
After satisfying the offline selections, the signal efficiency of the employed triggers is higher than 99\% (95\%) for the \EE (\MM) channel. Events are required to have at least one reconstructed
primary vertex with at least two associated tracks with \pt > 400~\MeV{}.
Among all the vertices in the event, the one with the highest
sum of squared transverse momenta of the associated tracks is
chosen as the primary vertex. Events containing \bjets are vetoed to reduce
contamination from top-quark background.

All selected events are required to have exactly two same-flavour
leptons ($ee,\mu\mu$) and two jets with $\pT > 100~\GeV{}$ and $|\eta| < 2.0$.
The main SM background after these requirements is the \Zjet process in the \OC and electron \SC channels,
while diboson production is the main background in the muon \SC channel.
The \Zjet process enters the \SC channel due to misidentification of electron charge.

Selected events are classified into orthogonal categories, called analysis regions, which serve different purposes. Signal regions
(SR) are designed to contain the majority of expected signal events and are used to extract the signal yields. Control regions (CR) and
validation regions (VR) are defined by reversing some of the SR event selection criteria and are used to constrain and validate
background predictions respectively. The analysis region selection criteria are summarised in Table~\ref{tab:AllRegions}.
To suppress background from \Zjet processes in the 
\SROSEE, \SROSMM, \SRSSEE and \SRSSMM signal regions,
the invariant masses of the two leptons (\mll) and the
two jets (\mjj) must be larger than 400~\GeV{} and 110~\GeV{}, respectively.
In addition, the scalar \pT sum of the selected two leptons (\ET for
electrons) and the two most energetic jets (\HT) must be larger than
400~\GeV{} to further reject the SM background, which exhibits lower
transverse energies than
the signal. Since the expected event yields in the \OC channel are larger
than those in the \SC channel, the CR and VR definitions differ slightly, as shown in
Table~\ref{tab:AllRegions}. The variable used to ensure the orthogonality between CRs, VRs and SRs is \mll, except in the \CROSEM which has the same 
kinematic selection as the SRs, but with different-flavour leptons. The \CROSEE and \CROSMM provide a high-purity sample of \Zjet events
and are used to constrain the yields from this process. Similarly, the \CROSEM is used to constrain events from top-quark processes. 
For the \SC channel, the main prompt SM background is diboson production and its yield is constrained in both the \CRSSEE and \CRSSMM.
The rate of \Zjet events, which enter the analysis regions only through charge misidentification, as described in Section~\ref{sec:background}, is constrained only in the \CRSSEE\footnote{The probability of muon charge misidentification is negligible because muon bremsstrahlung radiation is greatly reduced compared to electron, and
muon tracks are measured
in both the inner detector and the muon spectrometer, which provides a much larger lever arm for the curvature measurement.}. The invariant mass window used to define the \CRSSEE is reduced
to the interval $[110, 300]$ \GeV, to avoid overlap with the region around
the $Z$ mass peak, used to estimate the electron charge misidentification
probability. 

Once all selections are applied, the signal acceptance times efficiency -- evaluated with simulated signal events -- varies from
54\% in the (\WR,\NR) high-mass regime to $\sim30\%$ at the edge of the already excluded (\WR,\NR) low-mass region.

\begin{table}[htpb]
\caption{Summary of all regions defined in the analysis divided into the \OC and \SC channels. 
The table is split into two blocks:
the top half indicates the mass range of the dilepton final state, whereas the bottom half indicates
the event selection criteria used for a given region. The flavour combinations
are summarised as follows: $\ell^{\pm}\ell^{\mp}=\left\lbrace e^{\pm}e^{\mp},\mu^{\pm}\mu^{\mp}\right\rbrace$, $\ell^{\pm}\ell^{\pm}=\left\lbrace e^{\pm}e^{\pm},\mu^{\pm}\mu^{\pm}\right\rbrace$, $\ell^{\pm}\ell^{'\mp}=\left\lbrace e^{\pm}\mu^{\mp}\right\rbrace$.
Pairs of values $[X,Y]$ indicate the minimum and maximum values the quantity in question may take in the analysis region in question.
}
\label{tab:AllRegions}
\renewcommand{\arraystretch}{1.3}
\centering
\footnotesize
\setlength\tabcolsep{3.2pt}
\begin{adjustbox}{width=1\textwidth}
\begin{tabular}{l|ccc|cc|cc}
\toprule
Region 
 & \multicolumn{3}{c|}{Control region}
 & \multicolumn{2}{c|}{Validation region}
 & \multicolumn{2}{c}{Signal region} \\ \hline

Channel                                &  CR($\ell^{\pm}\ell^{\mp}$)  & CR($\ell^{\pm}\ell^{'\mp}$) & CR($\ell^{\pm}\ell^{\pm}$) & VR($\ell^{\pm}\ell^{\mp}$)&  VR($\ell^{\pm}\ell^{\pm}$)     &  SR($\ell^{\pm}\ell^{\mp}$) & SR($\ell^{\pm}\ell^{\pm}$)\\

\midrule

 $m_{ee}$ [\GeV]       & $[60,110]$ &  ---              & $[110,300]$   &$[110,400]$   & $[300,400]$&   $> 400$   &$> 400$\\

 $m_{\mu \mu}$ [\GeV] & $[60,110]$  & ---              & $[60,300]$   &  $[110,400]$   & $[300,400]$&   $> 400$    & $> 400$\\

 $m_{e \mu }$ [\GeV]   &   ---          & $> 400$ &     ---       &        ---        &    ---       &         ---            & --- \\    

 \midrule

 $H_{\text{T}}$ [\GeV] & $> 400$ & $> 400$ &   --- & $> 400$ & --- & $> 400$ & $> 400$\\

 $m_{jj}$ [\GeV]        & $> 110$ & $> 110$ &   --- & $> 110$ & --- & $> 110$ & $> 110$\\

 Jet \pt [\GeV]        & $>100$  & $>100$  & $>50$  & $>100$ & $>50$ & $>100$ & $>100$\\



 \bottomrule
\end{tabular}
\end{adjustbox}
\end{table}

\FloatBarrier

\section{Background estimation}
\label{sec:background}
The composition of the SM background is substantially different
in the \OC and \SC channels, requiring different background estimation
techniques in the two cases, although SM backgrounds containing a prompt
lepton (referred to as `prompt SM backgrounds' hereafter) are
estimated using simulations in both channels.
Prompt leptons are defined as leptons originating from $Z$, $W$, and $H$ boson decays,
or leptons from $\tau$ decays if the $\tau$-lepton originates from a prompt decay
(e.g. $Z\rightarrow\tau^+\tau^-$).
Unless otherwise stated, background processes are found to be well modelled by simulation.

\begin{figure}[htbp]
\centering
\includegraphics[width=0.45\textwidth]{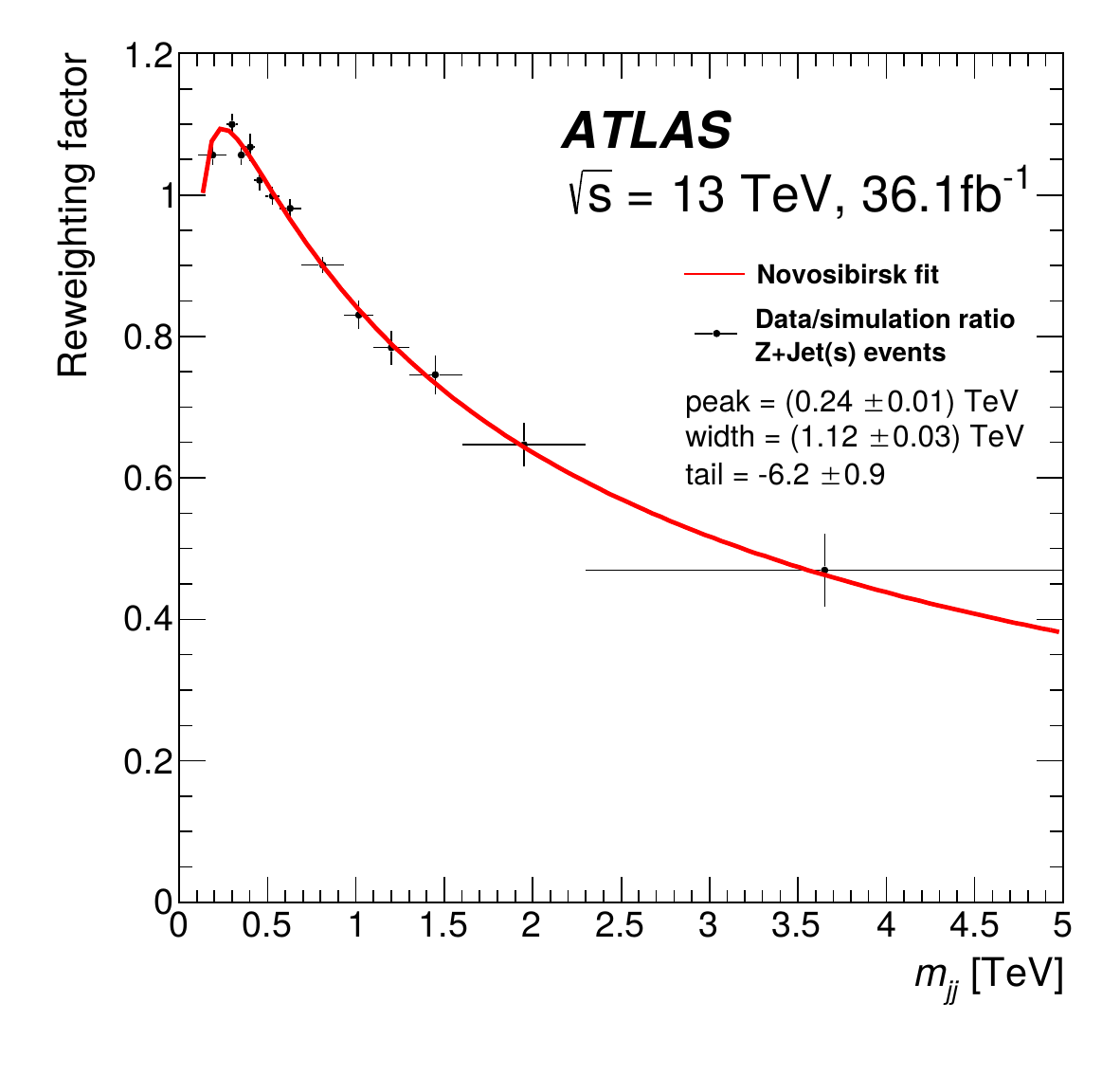}
\caption{Ratio between data and simulation for \Zjet events collected in \CROSEE and \CROSMM
as a function of the dijet invariant mass,
with the result of the Novosibirsk fit superimposed. The fitted values for
the peak and width in TeV, and tail (unitless) are shown.
The binning is adjusted to minimise statistical fluctuations with 36.1
fb$^{-1}$ of data.} 
\label{fig:m_jj_fit}
\end{figure}

The main prompt SM backgrounds contributing to the \OC channel
are from top-quark events (mainly \ttbar) and \Zjet production, with contributions of $\sim$49\% and $\sim$35\% respectively in the \SROSEE and $\sim$55\% and $\sim$37\% in the \SROSMM.
Minor contributions arise from diboson (mainly \WZ and \ZZ) and \Wjet events.
The \mjj spectrum of simulated \Zjet events is not correctly modelled by the simulation samples in
the \CROSEE and \CROSMM. The effect is not visible in the \SC channel
because \Zjet production is a less dominant background
and therefore the statistical uncertainty is much larger.
A data-driven procedure is applied to the simulated \Zjet
events to correct for this mismodelling.
A reweighting factor is derived from the regions \CROSEE and \CROSMM and applied to all \OC regions.
It ranges from $1.1$ in the low-\mjj region to $0.4$ in the
high-\mjj region above 3~\TeV{}.
The reweighting function is evaluated by fitting the ratio between data and simulation, after
subtracting the non-\Zjet contributions from both of them. 
The function is a Novosibirsk-type function with three free parameters (peak and width, measured in TeV, and tail, unitless)
and related to the weight by

\begin{subequations}
\label{eq:novosibirsk}
\begin{align}
  k_{1} & = \ln\bigg[1.0-\frac{(\mjj-{\rm peak})\cdot{\rm tail}}{{\rm width}}\bigg] ,\\
  k_{2} & = \frac{\sinh^{-1}(\sqrt{\ln4} \cdot {\rm tail})}{\sqrt{\ln4}} ,\\
 \mathrm{weight} & = \exp\bigg[-\frac{0.5}{k^2_{2}}\,k^2_{1}-0.5\,k^2_{2}\bigg] .
\end{align}
\end{subequations}
The resulting reweighting factor with the best fit function
is shown in Figure~\ref{fig:m_jj_fit}.

In the \SC channel, prompt SM processes such as \Zjet ($\sim$18\%) and diboson ($\sim$22\%) 
production contribute in the \SRSSEE. Diboson production also contributes significantly ($\sim$37\%) 
to the \SRSSMM. A small fraction of top-quark events is present in both \SRSSEE ($\sim$7\%) and \SRSSMM ($\sim$3\%). 
The largest contribution, $\sim$53\% in \SRSSEE and $\sim$60\% in \SRSSMM, is a reducible background arising
from events with misidentified or non-prompt electrons and muons, collectively called `fakes'.
A data-driven approach is used to assess the fake-lepton contributions as described in the following.

In the electron \SC channel, electron charge misidentification
is an additional source of SM background events mainly coming from \Zjet and top-quark production.
A discrepancy between data and simulation is observed in these events, and data-driven
correction factors are applied to simulated events. The charge misidentification of electrons is mainly due to bremsstrahlung radiation,
with subsequent photon conversion into an electron--positron pair.
The probability of charge misidentification is measured in data using
$Z \rightarrow e^+e^-$ events by comparing the numbers of \OC and \SC events. The probability
is parameterised in terms of the electron $\ET$ and $|\eta|$~\cite{EXOT-2016-07}.
It is measured in narrow $\ET$ bins with the last bin extending above \SI{250}{\GeV}.

The misidentified-lepton background is estimated with the `fake-factor' method,
implemented in the same way as in \Ref{\cite{EXOT-2016-07}}.
The implementation of the method requires two different lepton samples.
The first one is defined by the leptons entering any analysis region,
selected as in Table~\ref{tab:LepReco}, and called `tight' ($T$).
The second one is orthogonal to the tight selection and is
called `loose' ($L$). It is constructed by loosening some of the tight lepton
requirements and by requiring that these leptons fail to satisify the tight lepton criteria. 
The identification criteria for electrons are loosened to \texttt{LHLoose}~\cite{ATLAS-CONF-2016-024,ATL-PHYS-PUB-2015-041}
and the isolation criteria are not imposed. Similarly, loose muons have a relaxed impact
parameter requirement, $|d_0| / \sigma _{d_0} < 10$, and no isolation requirement.
The fake-factor ($F$), quantifying the probability for a misidentified lepton to be reconstructed as a prompt lepton,
is defined as the ratio of the number of tight to loose leptons $F = N_T / N_L$ and measured in dedicated `fake-enriched' regions. 

The regions, summarised in \Tab{\ref{tab:FakeEnrichedRegions}}, are enriched in fake leptons 
originating from dijet and \Wjet production.
They are selected using single-lepton triggers
and must satisfy low missing transverse momentum (\MET) requirements, to suppress \Wjet production
where a real lepton from a $W$ boson decay is present.
One charged lepton and at least one (two) jet(s) must be present in the
event in the muon (electron) channel. Furthermore, contamination from prompt leptons in fake-enriched regions
is subtracted using simulated events. This contamination is up to $40\%$ in the electron channel,
and in the muon channel for \pt ($<\SI{100}{\GeV}$), while it becomes dominant for high \pt muons.
The selection criteria for the electron fake-enriched
region are designed to mimic as much as possible the signal region selection in order to
ensure a similar composition of various sources of fake electrons: decays of
light-flavour hadrons into light leptons, photon conversions, and jets faking electrons. In the muon channel,
the major source of fake muons originates from decays of heavy-flavour hadrons and the muon fake-enriched 
region is optimised to maximise the number of such events.
The fake-factors are parameterised in terms of lepton flavour
($e$ or $\mu$), \ET (\pt) for electrons (muons), and $\eta$.
For electrons, they vary from 0.3 to 0.5 in the barrel region and from 0.4 to 0.9 in the endcap region. For muons, they are about 0.4 for $\pt < \SI{50}{\GeV}$ and rise to 1.0 for $\pt > \SI{50}{\GeV}$.

\begin{table}[htbp]
\caption{Selection criteria defining the dedicated fake-enriched regions, used to measure the electron and muon fake-factors as described in the text.}
\label{tab:FakeEnrichedRegions}
\renewcommand{\arraystretch}{1.3}
\centering
\small
\begin{tabular}{c|c}
\toprule
\multicolumn{2}{c}{Selection for fake-enriched regions}\tabularnewline
\multicolumn{1}{c}{Electrons} & Muons\tabularnewline
\midrule
Single-electron trigger & Single-muon trigger  \tabularnewline
$b$-jet veto & $b$-jet veto              \tabularnewline
One electron and at least two jets & One muon and one jet  \tabularnewline
$p_{\mathrm{T}}(\mathrm{jet})>\SI{50}{\GeV}$ & $p_{\mathrm{T}}(\mathrm{jet})>\SI{35}{\GeV}$ \tabularnewline
-- & $\Delta\phi(\mu,\mathrm{jet})>2.7$ \tabularnewline
$E_{\mathrm{T}}^{\mathrm{miss}}<\SI{25}{\GeV}$ & $E_{\mathrm{T}}^{\mathrm{miss}}<\SI{40}{\GeV}$ \tabularnewline
\bottomrule
\end{tabular}
\end{table}

Once the factor $F$ is measured, the contribution of the number of events with at least one
misidentified lepton in the analysis regions ($N^{\mathrm{fake}}$) is evaluated using:

\begin{linenomath}
\begin{equation}
N^{\mathrm{fake}} = \left[ F(N_{TL}^{\mathrm{data}}+N_{LT}^{\mathrm{data}}) -F^2N_{LL}^{\mathrm{data}}\right] - \left[ F(N_{TL}^{\mathrm{MC}}+N_{LT}^{\mathrm{MC}}) -F^2N_{LL}^{\mathrm{MC}}\right]^{\mathrm{prompt\,only}} .
\label{eq:FFformula}
\end{equation}
\end{linenomath}

In these regions the factor $F$ is applied to events containing at least one loose lepton,
$N_{TL},N_{LT}$ and $N_{LL}$, populating the `sideband' regions, namely $TL$, $LT$ and $LL$.
In these events, either the subleading lepton is loose ($TL$), or the leading lepton is loose ($LT$), or 
both leptons are loose ($LL$). 
The residual prompt-lepton components in the sideband regions are subtracted using simulated events.
The estimated backgrounds due to fakes and electron charge misidentification are validated in the
dedicated VRs defined in Table~\ref{tab:AllRegions}.

\FloatBarrier

\section{Systematic uncertainties}
\label{sec:syst}
Several sources of systematic uncertainty are considered.
They correspond to experimental and theoretical sources affecting both the background and signal predictions.
All the sources of systematic uncertainty considered here affect the
total event yield. They also alter the distributions of the
variables used in the statistical analysis (Section~\ref{sec:result}) with
the exception of the uncertainties in the luminosity and cross-sections.

The uncertainty in the 2015 and 2016 integrated luminosity is $2.1\%$,
derived with the methodology described in Ref.~\cite{DAPR-2013-01},
and using the LUCID-2 detector for the baseline luminosity measurements \cite{Avoni:2018iuv},
from a calibration of the luminosity scale using
$x$--$y$ beam-separation scans.

A set of experimental systematic uncertainties arise from
the calibrations of the lepton and jet energy and momentum, the lepton detection
efficiencies and isolation, and the trigger efficiency. The largest uncertainty in the total
SM yield arises from the energy calibration and smearing of jets, derived in Ref.~\cite{PERF-2016-04},
and is between 5\% and 10\% depending on the invariant mass of the \lljj system.
Experimental systematic uncertainties associated with lepton reconstruction,
identification, isolation and trigger efficiencies, as well as energy or momentum
calibration and $b$-jet tagging, vary between 0.5\% and 4\% of the total SM yield.

There are two additional experimental sources of systematic uncertainty
in the \SC channel due to the data-driven background estimation techniques.
The uncertainty related to the charge misidentification probability of
electrons arises from the statistical uncertainty of both the data and the simulated
samples of \Zjet events used to measure this probability. The
uncertainty ranges between $10\%$ and $20\%$ as a function of the electron
\ET and $\eta$. Next, the uncertainty on the fake estimation arises from
the unknown composition of fakes, 
as well as from statistical uncertainty and prompt lepton subtraction used
to derive $F$ in the fake-enriched regions. The uncertainty
due to the composition of fakes is estimated by varying the criteria to select the nominal
sample for the fake-factor measurement (\Tab{\ref{tab:FakeEnrichedRegions}}). 
In the electron channel, the \MET requirement is increased to $<$\SI{100}{\GeV},
the two-jet requirement is dropped and one jet recoiling against the electron 
(analogous to the nominal muon fake-factor measurement) is required. 
In the muon channel, the nominal $|d_0| / \sigma _{d_0}$ and \MET requirements
are varied up and down respectively by one unit and \SI{10}{GeV}. Furthermore, 
the uncertainty in the yield of prompt leptons from $W$ and $Z$ boson decays is estimated by varying
the total SM prediction of the simulated samples by $\pm 10\%$ in the muon channel and
$\pm 30\%$ in the electron channel. These values represent the size of the uncertainty due to the choice of
QCD renormalisation (\muR) and factorisation (\muF) scales, \alphaS, and PDF uncertainties.
The different size of the uncertainty between channels 
arises from the different fake-enriched region definitions (Table~\ref{tab:FakeEnrichedRegions}), 
requiring exactly one jet and at least two jets for the muon and electron channels respectively.
The total uncertainty in the fake-factors 
ranges between $10\%$ and $50\%$ across all $\ET(e, \mu)$ and $\eta$ bins for both flavours.

In the \OC channel, an additional uncertainty is taken into account for the \mjj reweighting of the \Zjet process.
This is evaluated by comparing the shape difference between the simulated and reweighted \mjj distribution and that
measured in data, using the corresponding OS VR: the uncertainty in the reweighting factor is
found to be between 5\% and 20\%, depending on the dijet invariant mass.

The theory uncertainties estimated for the \Zjet and diboson background processes include the choice of QCD renormalisation (\muR)
and factorisation (\muF) scales, choice of the PDF set, \alphaS, and PDF uncertainty.
The QCD scale uncertainty is estimated by varying \muR and \muF to half and twice their nominal values.
The PDF uncertainty is estimated using the envelope of the NNPDF3.0 PDF set, as recommended in Ref.~\cite{Botje:2011sn}.
In addition, the MMHT2014~\cite{Harland-Lang:2014zoa} and CT14NNLO~\cite{dulat2015ct14} PDF sets
are used to estimate the uncertainty due to the PDF choice.
Moreover, the uncertainty due to \alphaS is evaluated by varying its
nominal value of 0.118 by $\pm 0.001$. The largest theory uncertainty generally originates from the QCD scales variations,
and is between 20\% and 40\%{}, depending on the simulated process and the invariant mass of the \lljj system.

The theory uncertainties estimated for \ttbar processes are as follows.
The uncertainty from hard-scatter generation is evaluated by comparing the \POWHEGBOX and \MGaMC generators, both interfaced to the \PythiaEight parton shower model, as recommended in Ref.~\cite{Botje:2011sn}.
The uncertainty due to the hadronisation and fragmentation model is determined by comparing the nominal \POWHEGBOX+ \PythiaEight
generated sample with the one generated by \POWHEGBOX interfaced to
\HerwigSeven~\cite{Bahr:2008pv} (version 7.0.1).
The uncertainty related to the amount of initial- and final-state radiation is assessed by varying parton shower settings.
The largest theory uncertainty generally originates from the amount
of initial- and final-state radiation and is between 2\% and 5\% of the \ttbar process yield.

Finally, the theory uncertainty of the signal efficiency times acceptance amounts to 10\%. It is evaluated by varying
renormalisation and factorisation scales as described above and by using alternative PDF sets,
CTEQ6~\cite{Pumplin:2002vw} and MSTW~\cite{Martin:2009iq}. The \alphaS emission scale factor is also varied to half and twice the nominal value.
The uncertainty is dominated by the variation in factorisation scale.
The variations are performed using \textsc{SysCalc}~\cite{Kalogeropoulos:2018cke}.

\FloatBarrier

\section{Statistical analysis and results}
\label{sec:result}
The statistical analysis package \textsc{HistFitter}~\cite{Baak:2014wma} is used to 
implement a binned maximum-likelihood fit of the data distributions 
in the control and signal regions for the evaluation of 
the numbers of signal and background events.
For the scenario in which the \NR neutrino is a Majorana particle, the \OC and \SC channels are fitted simultaneously,
whereas for the Dirac neutrino, only the \OC channel is used in the fit.
The analysis regions are optimised for the high-mass regime,
where the CRs and the VRs have negligible signal contribution.
On the other hand, it is found that lower mass points yield some signal events in the CRs. 
This effect is accounted for in the statistical analysis when evaluating the signal strength
by simultaneously fitting the CRs and SRs. The VRs are not used in the fit,
but are only employed to confirm the validity of the background modelling
before the fit is carried out in the SRs.

In the \OC channel, different distributions are used in the likelihood fit depending on the mass
hierarchy of the signal sample. 
The \mlljj distribution is used as the discriminant for
the $\mWR > \mNR$ hypothesis, and the \mjj spectrum is used as the
discriminant for the $\mWR < \mNR$ hypothesis\footnote{For the \CROSEE and
  \CROSMM, the event yields are integrated into a single-bin and then fed
  into the fit.}.
In both cases, the same discriminant
variable is used in the control region and the signal region. 
In the \SC channel, \mjj is used in the CRs and \HT is used in the SR, 
regardless of the mass hierarchy, as it yields a better
separation between the signal and the background.
Additional mass hypotheses are added by interpolating the discriminating distributions obtained for existing signal simulation samples using a moment morphing technique \cite{Baak:2014fta}.

The likelihood is the product of a Poisson probability density function describing
the observed number of events, and Gaussian distributions to constrain
the nuisance parameters associated with the systematic uncertainties. The widths of the Gaussian distributions 
correspond to the magnitudes of these uncertainties, whereas Poisson distributions are used 
for the simulation statistical uncertainties. Additional free parameters are introduced to normalise the contributions
of \Zjet, diboson, and top-quark backgrounds to the data in the analysis control regions.
The fitted normalisation parameters are applied simultaneously in SRs.
The fitted normalisations are consistent with their SM predictions within the uncertainties.

\begin{table}[htpb]
\caption{Summary of the control regions used to fit the yields of the dominant SM background predictions.
CRs used to fit a certain SM prediction yield are marked with a check-mark (\ding{51}) and
CRs not used for this SM prediction are marked with a dash (-). }
\label{tab:fittedYields}
\centering
\begin{tabular}{l|ccccc}
\toprule
SM background yield   & \CROSEE      & \CROSEM      & \CROSMM   & \CRSSEE      & \CRSSMM      \\

\midrule 

OS \Zjet         & \ding{51}  & -          & \ding{51}  & -          & -          \\
SS \Zjet         & -          & -          & -          & \ding{51}  & -          \\
OS Top           & -          & \ding{51}  & -          & -          & -          \\
SS Diboson       & -          & -          & -          & \ding{51}  & \ding{51}  \\

\bottomrule
\end{tabular}
\end{table}

\begin{table}
\begin{center}
{\footnotesize
\caption{Numbers of expected background events in control and validation regions after the background-only fit in the opposite-charge channel, compared to the data. 
The quoted uncertainties correspond to the total uncertainty in the predicted event yield including correlations between the various sources of systematic uncertainties. Due to rounding the total background can differ from the sums of components.
Top-quark and \Zjet yields are floating in the fit.}
\label{tab:analysisRegionsOS_CRVR}
\begin{tabular}{ll @{\,}@{$\pm$}@{\,} r l @{\,}@{$\pm$}@{\,} r l @{\,}@{$\pm$}@{\,} r l @{\,}@{$\pm$}@{\,} r l @{\,}@{$\pm$}@{\,} r}
\toprule
           & \multicolumn{2}{c}{CR ($e^{\pm}e^{\mp}$)}            & \multicolumn{2}{c}{CR ($e^{\pm}\mu^{\mp}$)}            & \multicolumn{2}{c}{CR ($\mu^{\pm}\mu^{\mp}$)}            & \multicolumn{2}{c}{VR ($\mu^{\pm}\mu^{\mp}$)}            & \multicolumn{2}{c}{VR ($\mu^{\pm}\mu^{\mp}$)}              \\
\midrule
Data          & \multicolumn{2}{l}{29178}              & \multicolumn{2}{l}{201}              & \multicolumn{2}{l}{37378}              & \multicolumn{2}{l}{2794}              & \multicolumn{2}{l}{3327}                    \\
\midrule
Total background         & 29200 & 170  & 202 & 14   & 37360 & 190  & 2700 & 130    & 3290 & 140      \\
\midrule
        \Zjet         & 27900 & 190  & 1.2 & 0.1 & 35790 & 220 & 1306 & 28    & 1729 & 36      \\
        $t\bar{t}$, single-$t$, $t\bar{t}V$         & 524 & 71  & 162 & 14  & 628 & 88 &1250 & 130  & 1400 & 140      \\
        Diboson and W+jets         & 775 & 28  & 39.1 & 2.0  & 937 & 35 & 149.1 & 6.5  & 160.9 & 5.9      \\
 \bottomrule
\end{tabular}
}
\end{center}
\end{table}

\begin{table}
\begin{center}
{\footnotesize
\caption{Numbers of expected background events in signal regions after the background-only fit in the opposite-charge channel, compared to the data. 
The quoted uncertainties correspond to the total uncertainty in the predicted event yield including correlations between the various sources of systematic uncertainties. Due to rounding the total background can differ from the sums of components.
Top-quark and \Zjet yields are floating in the fit.}
\label{tab:analysisRegionsOS_SR}
%
\begin{tabular}{l l @{\,}@{$\pm$}@{\,} r l @{\,}@{$\pm$}@{\,} r}
\toprule
           & \multicolumn{2}{c}{SR ($e^{\pm}e^{\mp}$)}            & \multicolumn{2}{c}{SR ($\mu^{\pm}\mu^{\mp}$)}              \\
\midrule
Data          & \multicolumn{2}{l}{156}              & \multicolumn{2}{l}{169}                    \\
\midrule
Total background         & 152.2 & 8.4  & 165.9 & 8.9      \\
\midrule
        \Zjet         & 53.9 & 5.1  & 61.8 & 4.8      \\
        $t\bar{t}$, single-$t$, $t\bar{t}V$         & 74.0 & 7.6  & 89.6 & 8.3      \\
        Diboson and \Wjet         & 24.5 & 1.6  & 12.8 & 0.7      \\
 \bottomrule
\end{tabular}
}
\end{center}
\end{table}

\begin{table}
\begin{center}
{\footnotesize
\caption{Numbers of expected background events in analysis regions after the background-only fit in the same-charge channel, compared to the data.
The quoted uncertainties correspond to the total uncertainty in the predicted event yield including correlations between the various sources of systematic uncertainties. Due to rounding the total background can differ from the sums of components. Diboson and \Zjet yields are floating in the fit. Background processes with a negligible yield are marked with the dash (--).}
\label{tab:analysisRegions}
\begin{tabular}{ll @{\,}@{$\pm$}@{\,} r l @{\,}@{$\pm$}@{\,} r l @{\,}@{$\pm$}@{\,} r l @{\,}@{$\pm$}@{\,} r l @{\,}@{$\pm$}@{\,} r l @{\,}@{$\pm$}@{\,} r}
\toprule
           & \multicolumn{2}{c}{CR ($e^{\pm}e^{\pm}$)}            & \multicolumn{2}{c}{CR ($\mu^{\pm}\mu^{\pm}$)}            & \multicolumn{2}{c}{VR ($e^{\pm}e^{\pm}$)}            & \multicolumn{2}{c}{VR ($\mu^{\pm}\mu^{\pm}$)}            & \multicolumn{2}{c}{SR ($e^{\pm}e^{\pm}$)}            & \multicolumn{2}{c}{SR ($\mu^{\pm}\mu^{\pm}$)}              \\
\midrule
Data          & \multicolumn{2}{l}{304}              & \multicolumn{2}{l}{119}              & \multicolumn{2}{l}{33}              & \multicolumn{2}{l}{11}              & \multicolumn{2}{l}{11}              & \multicolumn{2}{l}{5}                    \\
\midrule
Total background         & 306 & 17  & 119 & 11  & 31.1 & 5.5  & 10.3 & 2.4  & 11.2 & 2.0  & 5.5 & 1.7      \\
\midrule
        \Zjet         & 100 & 31  & \multicolumn{2}{c}{--}  & 4.3 & 2.0  & \multicolumn{2}{c}{--}  & 2.0 & 0.9  & \multicolumn{2}{c}{--}      \\
        Fakes         & 119 & 23  & 40.4 & 9.6  & 14.1 & 4.6  & 4.1 & 2.0  & 5.9 & 1.9  & 3.3 & 1.7     \\
        Diboson         & 61 & 13  & 74 & 14  & 7.3 & 1.6  & 5.8 & 1.4  & 2.6 & 0.6  & 2.0 & 0.5      \\
        $t\bar{t}$, single-$t$, $t\bar{t}V$    & 25.8 & 5.9  & 4.3 & 0.8  & 5.4 & 1.7  & 0.35 & 0.08  & 0.8 & 0.3  & 0.14 & 0.05      \\
 \bottomrule
\end{tabular}
}
\end{center}
\end{table}

\begin{figure}[htpb]
\centering
\includegraphics[width=1.03\textwidth]{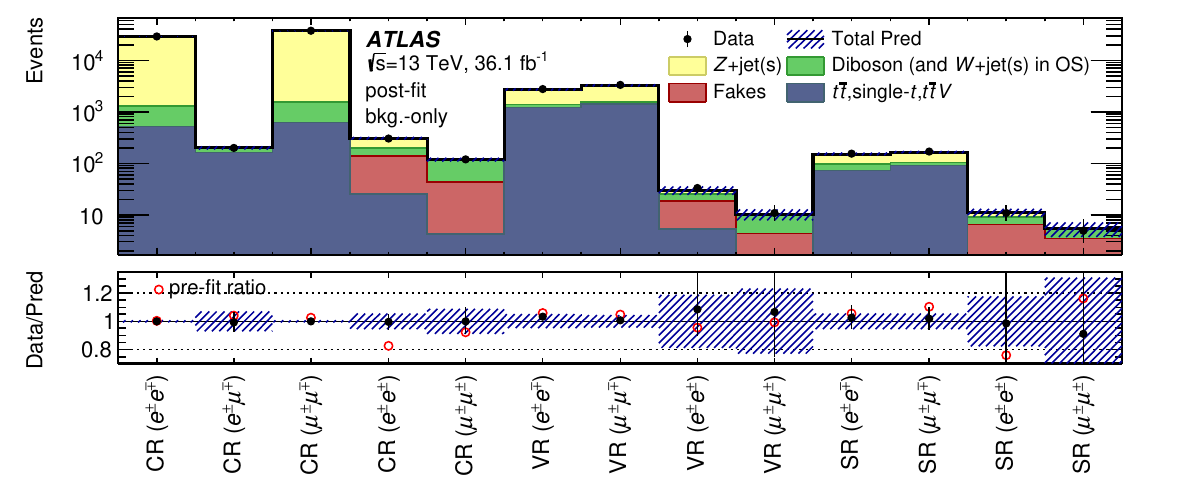}
\caption{
Number of observed and expected events in the control (CR),
validation (VR), and signal (SR) regions for all channels considered after the background-only fit.
The background expectation is the result of the fit described in the text using both \OC and \SC control regions.
The hatched bands include all post-fit systematic uncertainties. The correlation between sources of systematic uncertainties are taken into account. The pre-fit ratio is indicated with the red
hollow markers in the ratio plot. }
\label{fig:allAnalysisRegions}
\end{figure}

Due to the different object definitions and background components in the \OC and \SC channels, separate nuisance parameters
are used for all sources of uncertainties and fitted yields for the largest backgrounds, as summarised in Table~\ref{tab:fittedYields}.
Observed and predicted event yields and the corresponding uncertainties of the dominant backgrounds in all analysis regions
are presented in
Tables~\ref{tab:analysisRegionsOS_CRVR},~\ref{tab:analysisRegionsOS_SR}
and~\ref{tab:analysisRegions}. The information on observed and expected yields, along with their ratio for pre- (black filled markers) and post-fit (red hollow markers) yields, is presented in Figure~\ref{fig:allAnalysisRegions} in all the analysis regions. 
The \CROSEE and \CROSMM are used to extract the yield of \Zjet background, as the purity is above 95\%.
The \CROSEM is mostly populated by top-quark background events (80\% purity) and used to extract their normalisation.
In the \SC channel, the \Zjet contribution is estimated to be 
$\sim$33\%, determined in a fit using the \CRSSEE only, and with the diboson normalisation factor free to vary. 
The data-driven fake background normalisation is fixed,
which allows for the measurement of the \Zjet normalisation despite its low
contribution in the control region.

\begin{figure}[htbp]
\centering
\subfloat[]{
\includegraphics[width=0.48\textwidth]{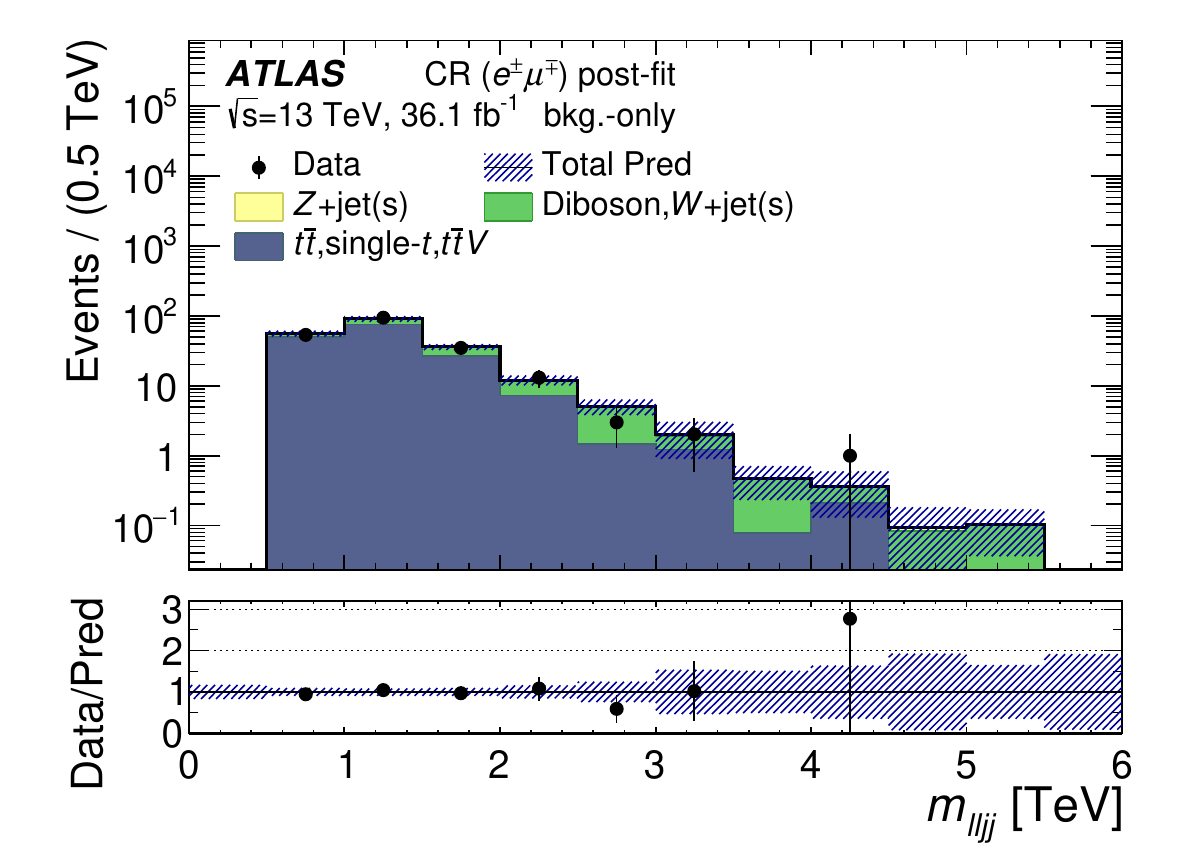}
\label{fig:histo_SR_emos_post}
}
\subfloat[]{
\includegraphics[width=0.48\textwidth]{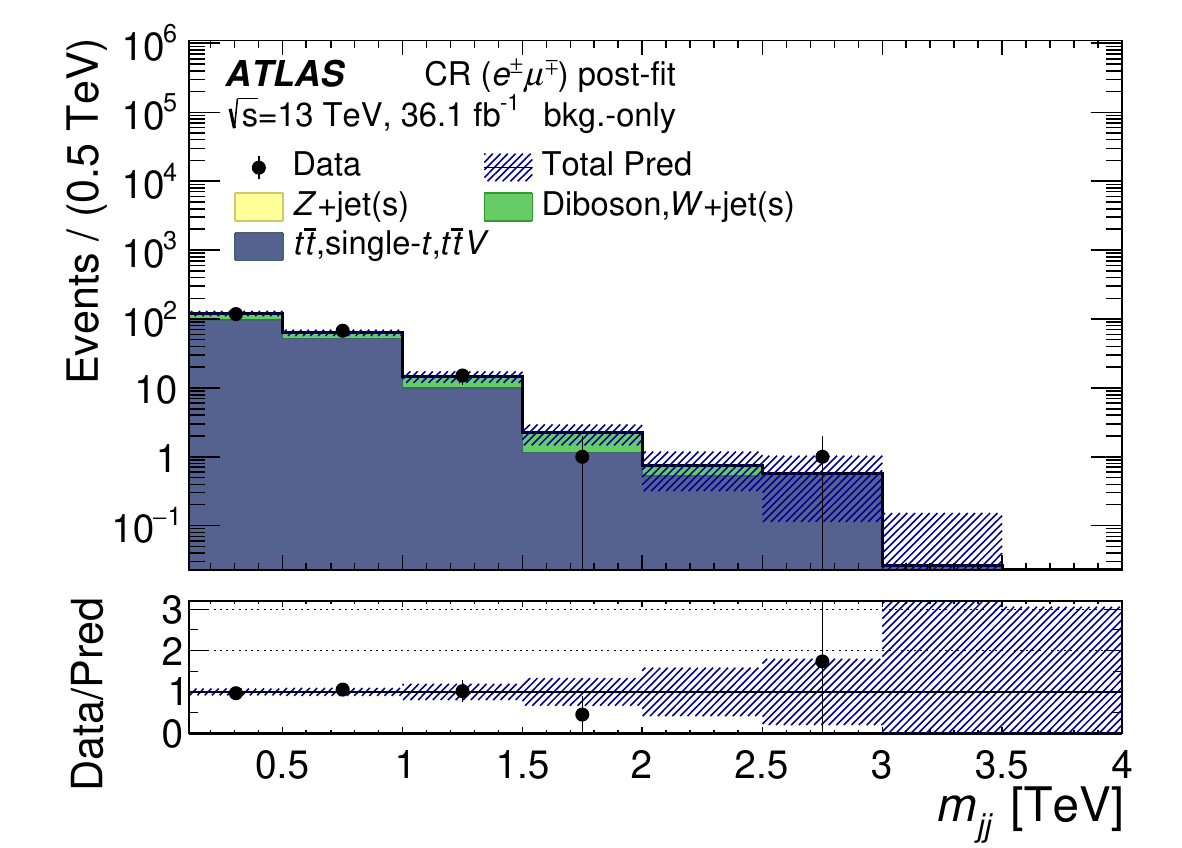}
\label{fig:histo_SR_emos_mjj_post}
}
\\
\subfloat[]{
\includegraphics[width=0.48\textwidth]{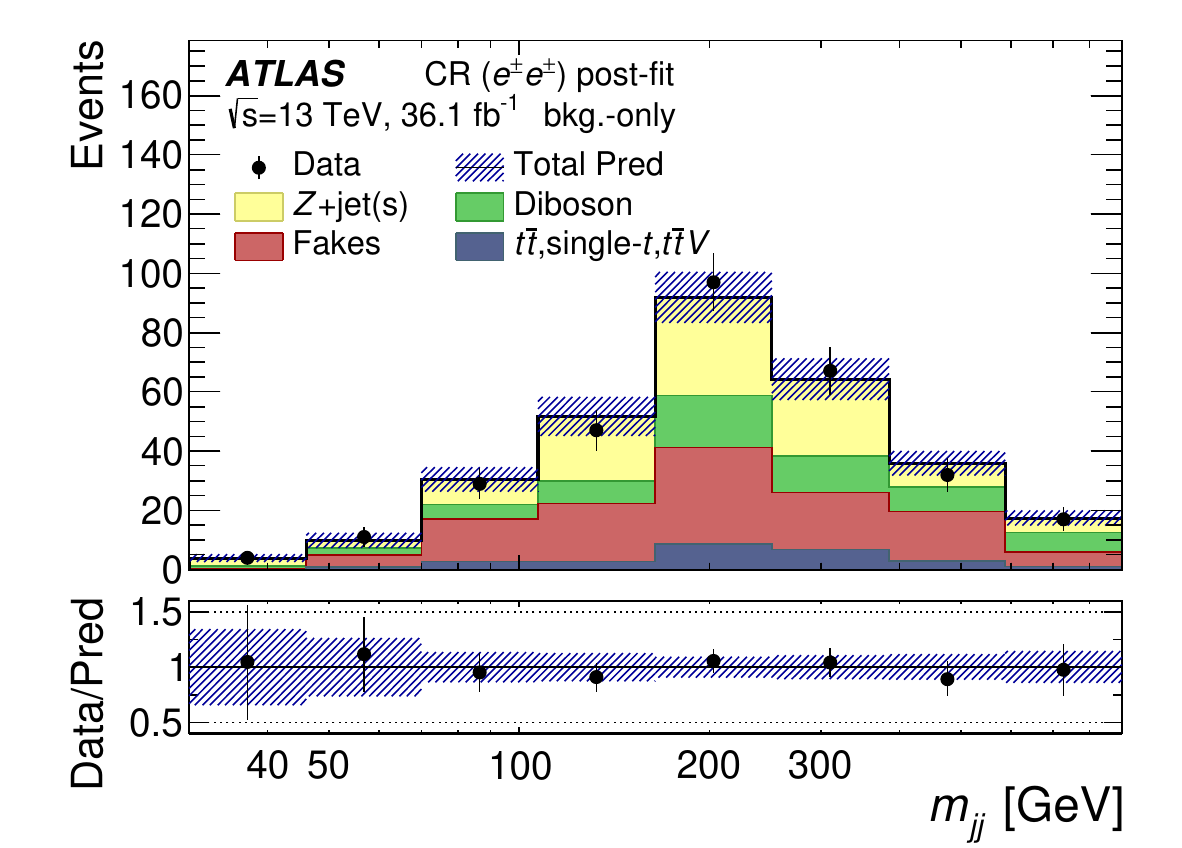}
\label{fig:histo_SSZCRee_post}
}
\subfloat[]{
\includegraphics[width= 0.48\textwidth]{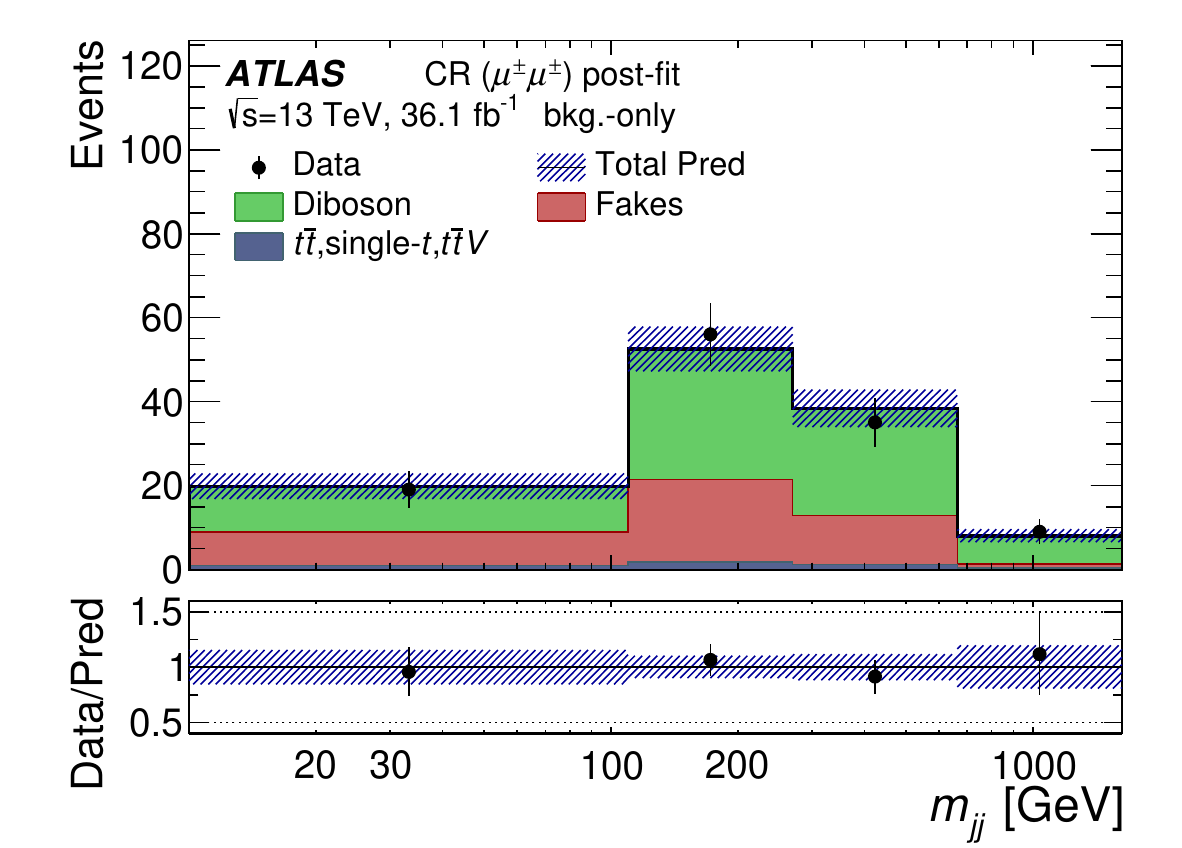}
\label{fig:histo_SSZCRmm_post}
}
\caption{Distributions for data and background predictions for
  discriminant variables in various control regions and channels after the background-only fit:
\protect\subref{fig:histo_SR_emos_post} \mlljj in \CROSEM,
\protect\subref{fig:histo_SR_emos_mjj_post} \mjj in \CROSEM,
\protect\subref{fig:histo_SSZCRee_post} \mjj in \CRSSEE, and
\protect\subref{fig:histo_SSZCRmm_post} \mjj in \CRSSMM.
The hatched bands include all post-fit systematic uncertainties. The correlation between sources of systematic uncertainties are taken into account. The last bin in each histogram contains the overflow.}
\label{fig:controlRegions_bkg}
\end{figure}

\begin{figure}[htbp]
\centering
\subfloat[]{
\includegraphics[width=0.48\textwidth]{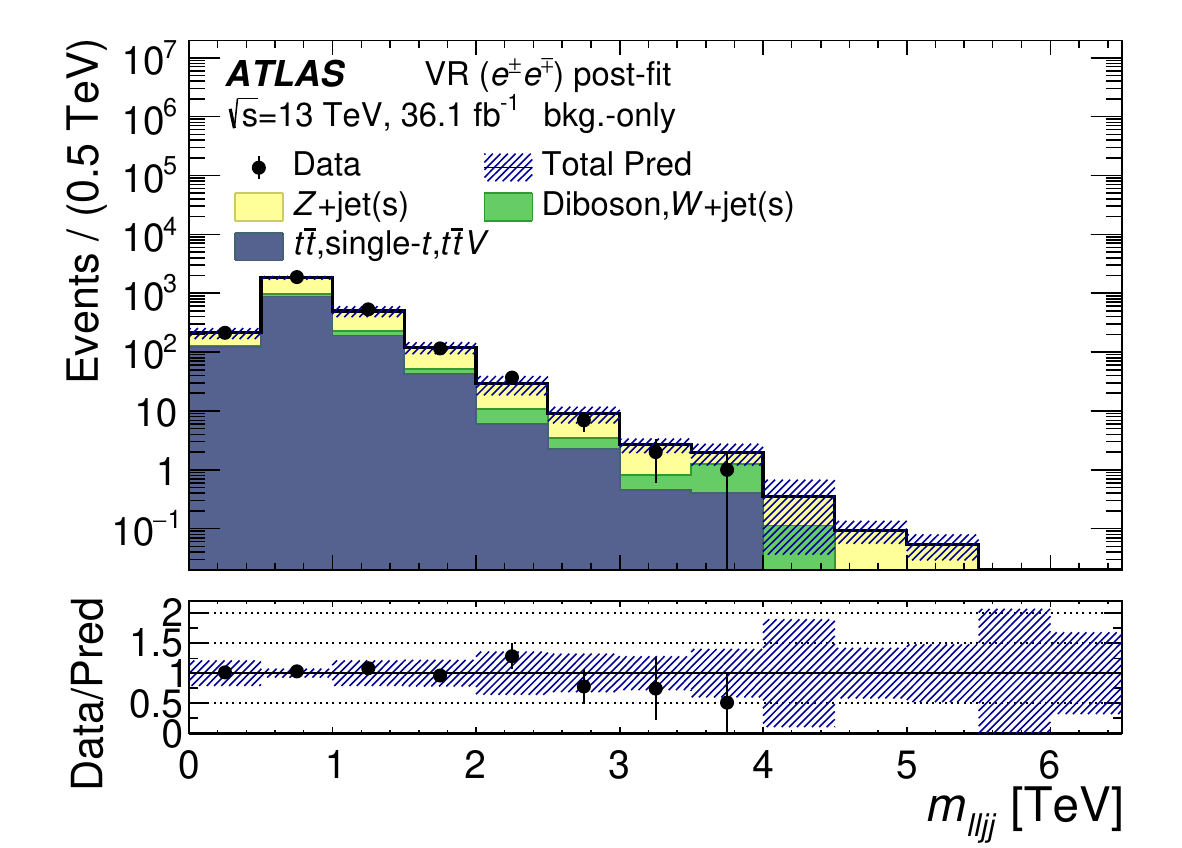}
\label{fig:histo_ZVR_eeos_post}
}
\subfloat[]{
\includegraphics[width= 0.48\textwidth]{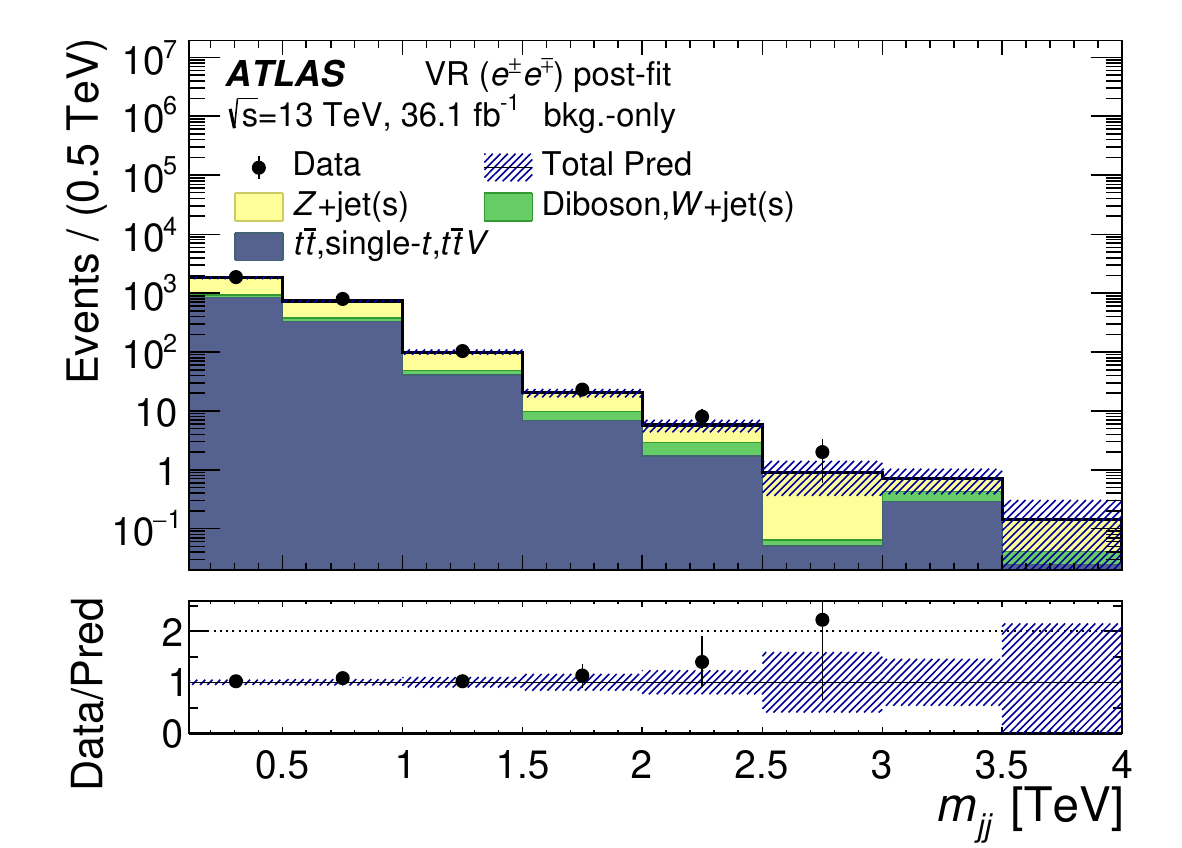}
\label{fig:histo_ZVR_eeos_mjj_post}
}
\\
\subfloat[]{
\includegraphics[width=0.48\textwidth]{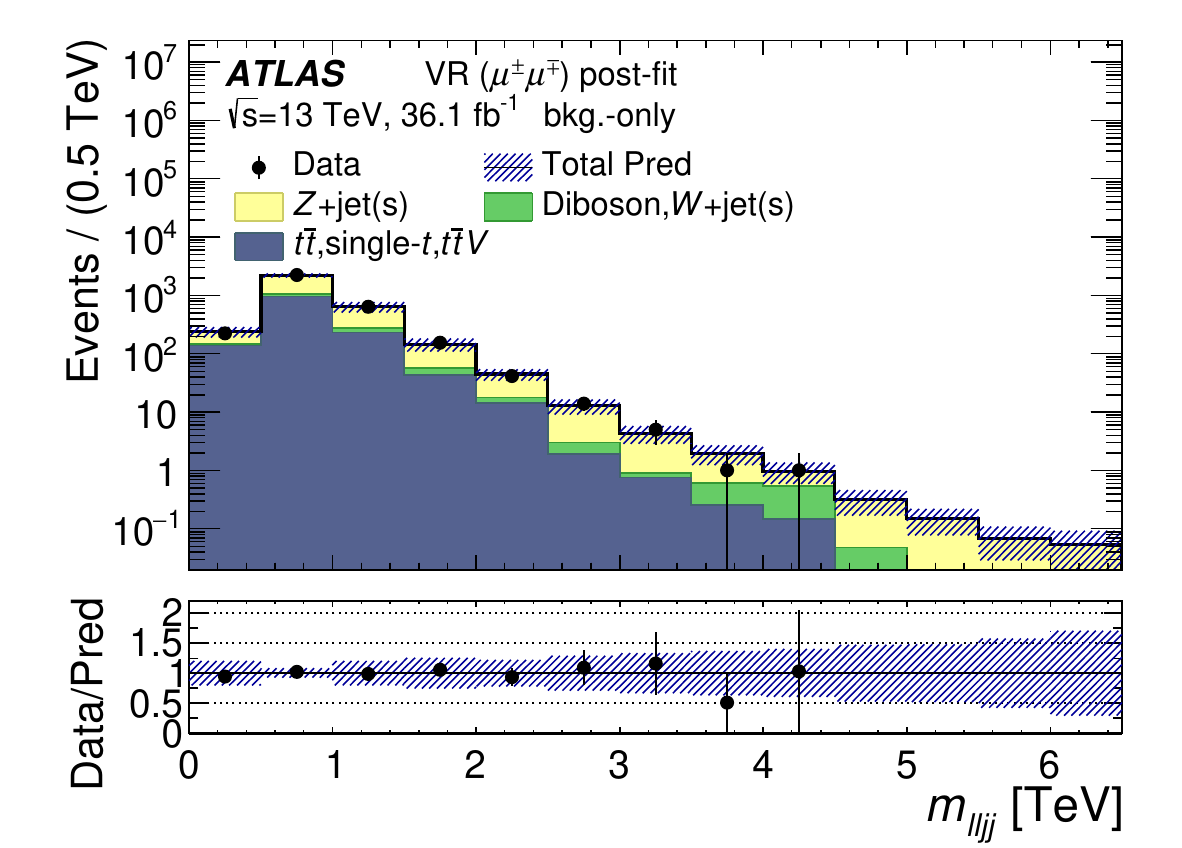}
\label{fig:histo_ZVR_mmos_post}
}
\subfloat[]{
\includegraphics[width= 0.48\textwidth]{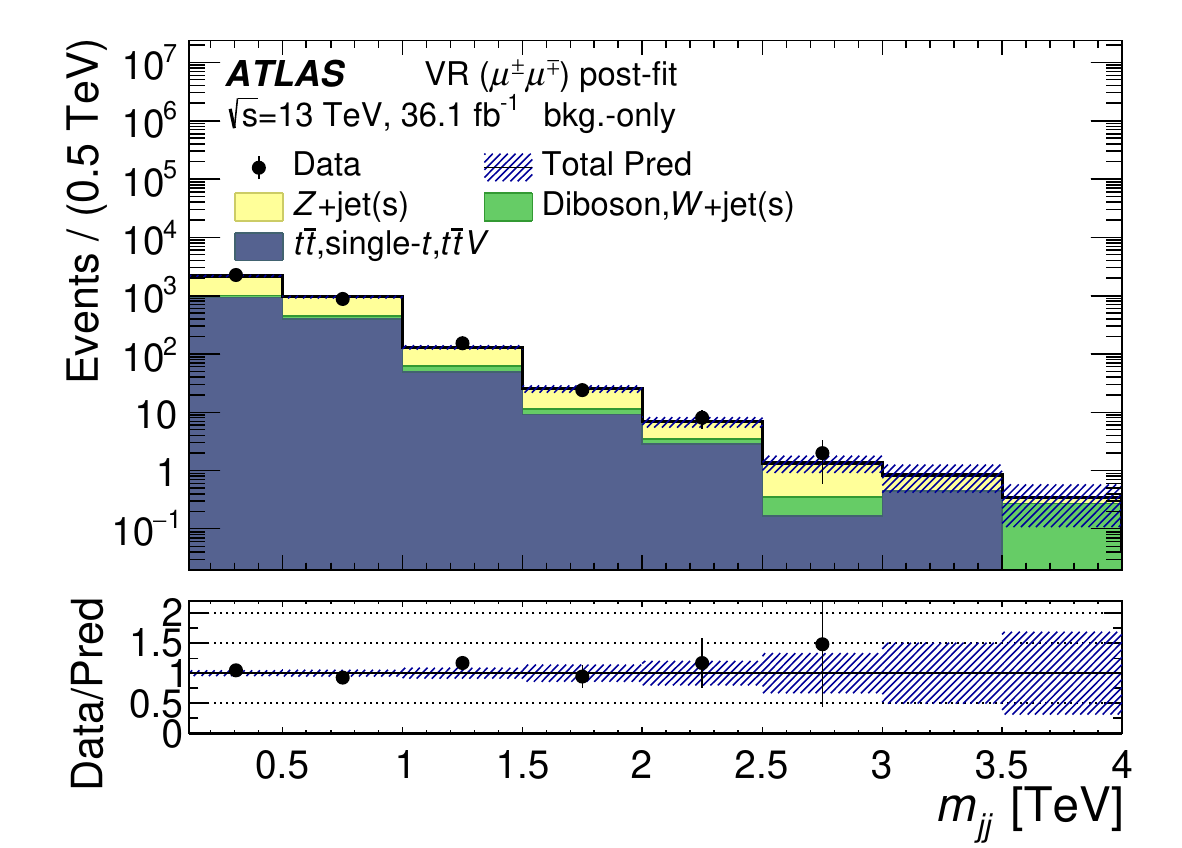}
\label{fig:histo_ZVR_mmos_mjj_post}
}
\\
\subfloat[]{
\includegraphics[width=0.48\textwidth]{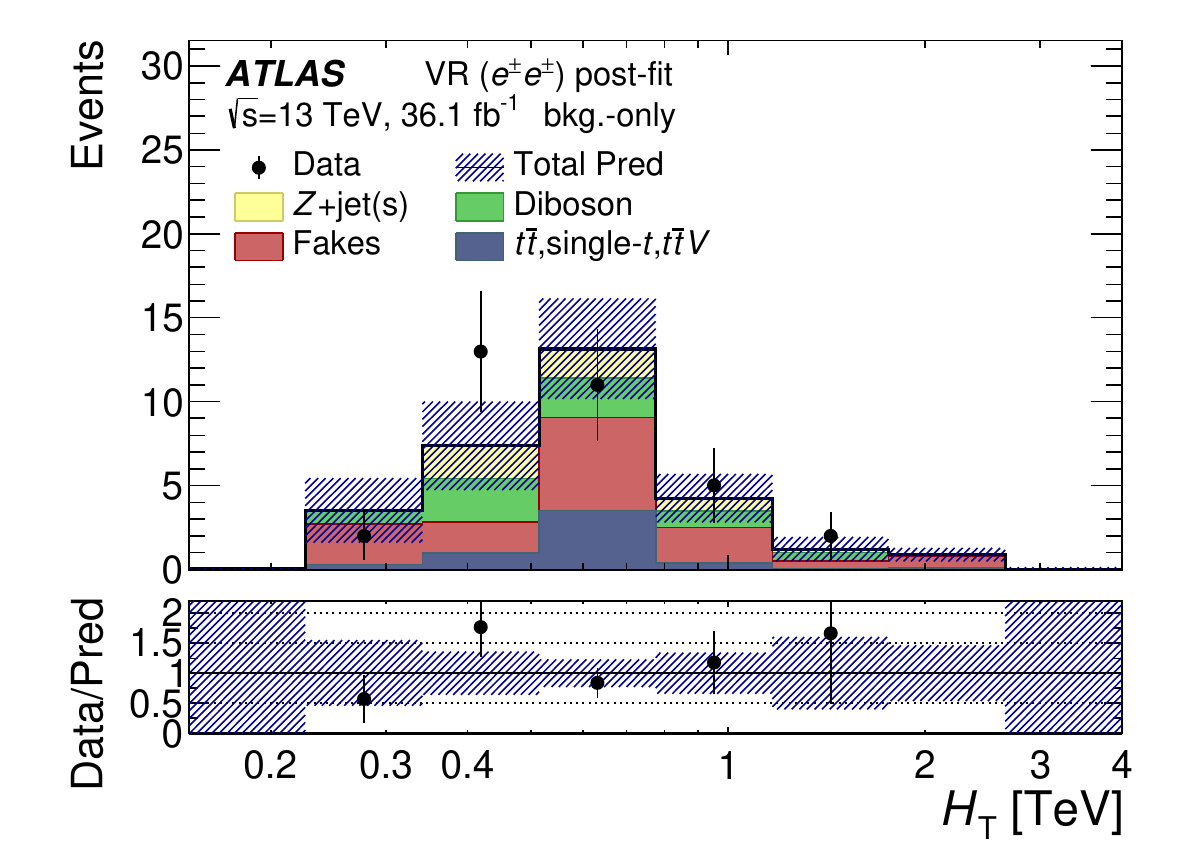}
\label{fig:histo_SSZVRee_post}
}
\subfloat[]{
\includegraphics[width= 0.48\textwidth]{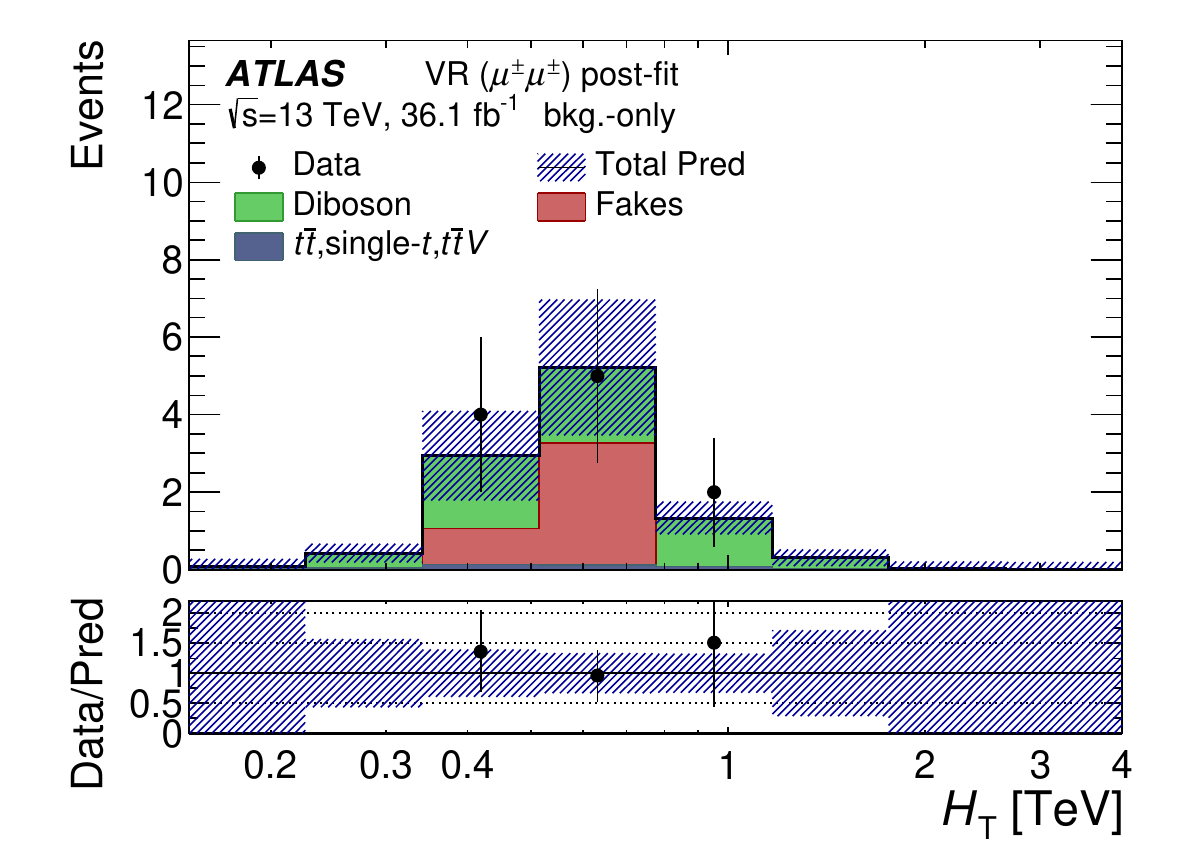}
\label{fig:histo_SSZVRmm_post}
}
\caption{Distributions for data and background predictions for
  discriminant variables in various validation regions and channels after the background-only fit:
\protect\subref{fig:histo_ZVR_eeos_post} \mlljj in \VROSEE,
\protect\subref{fig:histo_ZVR_eeos_mjj_post} \mjj in \VROSEE,
\protect\subref{fig:histo_ZVR_mmos_post} \mlljj in \VROSMM,
\protect\subref{fig:histo_ZVR_mmos_mjj_post} \mjj in \VROSMM,
\protect\subref{fig:histo_SSZVRee_post} \HT in \VRSSEE, and
\protect\subref{fig:histo_SSZVRmm_post} \HT in \VRSSMM.
The hatched bands include all post-fit systematic uncertainties. The correlation between sources of systematic uncertainties are taken into account. The last bin in each histogram contains the overflow.}
\label{fig:validationRegions_bkg}
\end{figure}

\begin{figure}[htbp]
\centering
\subfloat[]{
\includegraphics[width=0.48\textwidth]{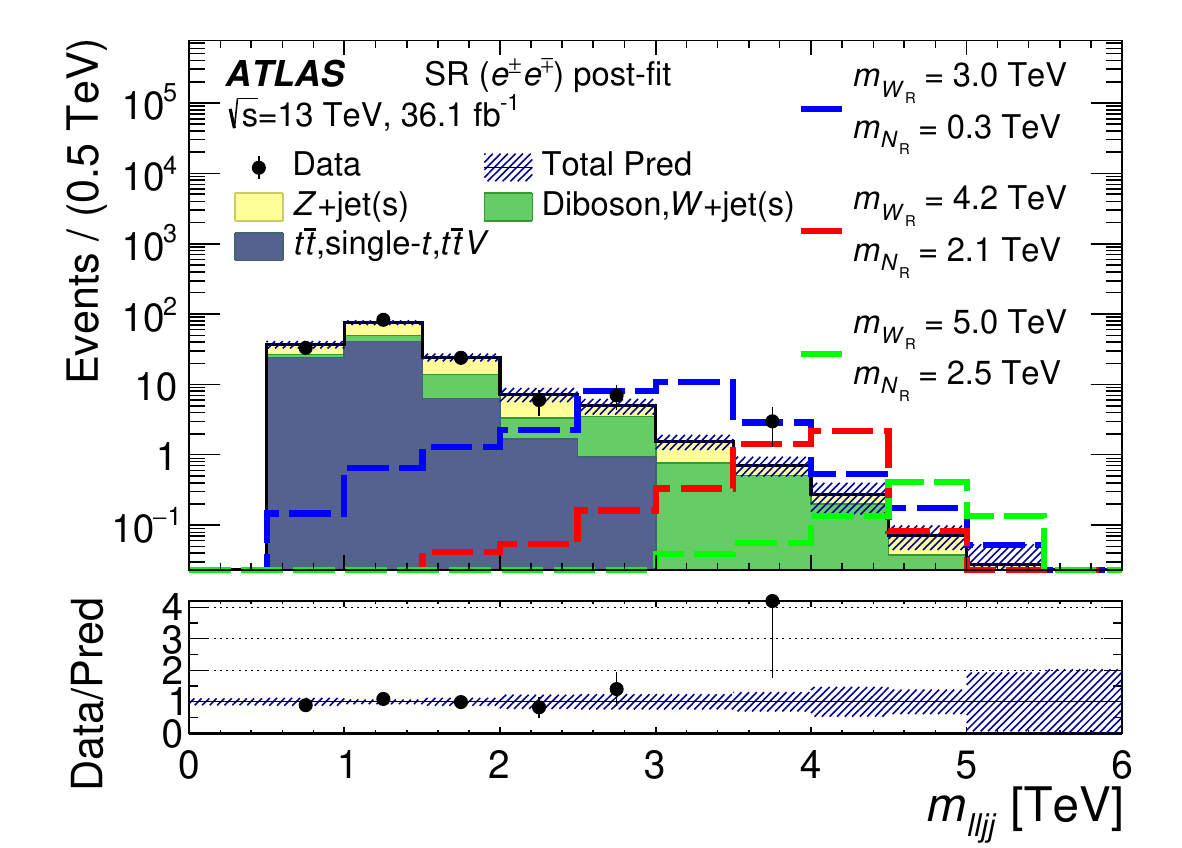}
\label{fig:histo_SR_eeos_excl_post}
}
\subfloat[]{
\includegraphics[width= 0.48\textwidth]{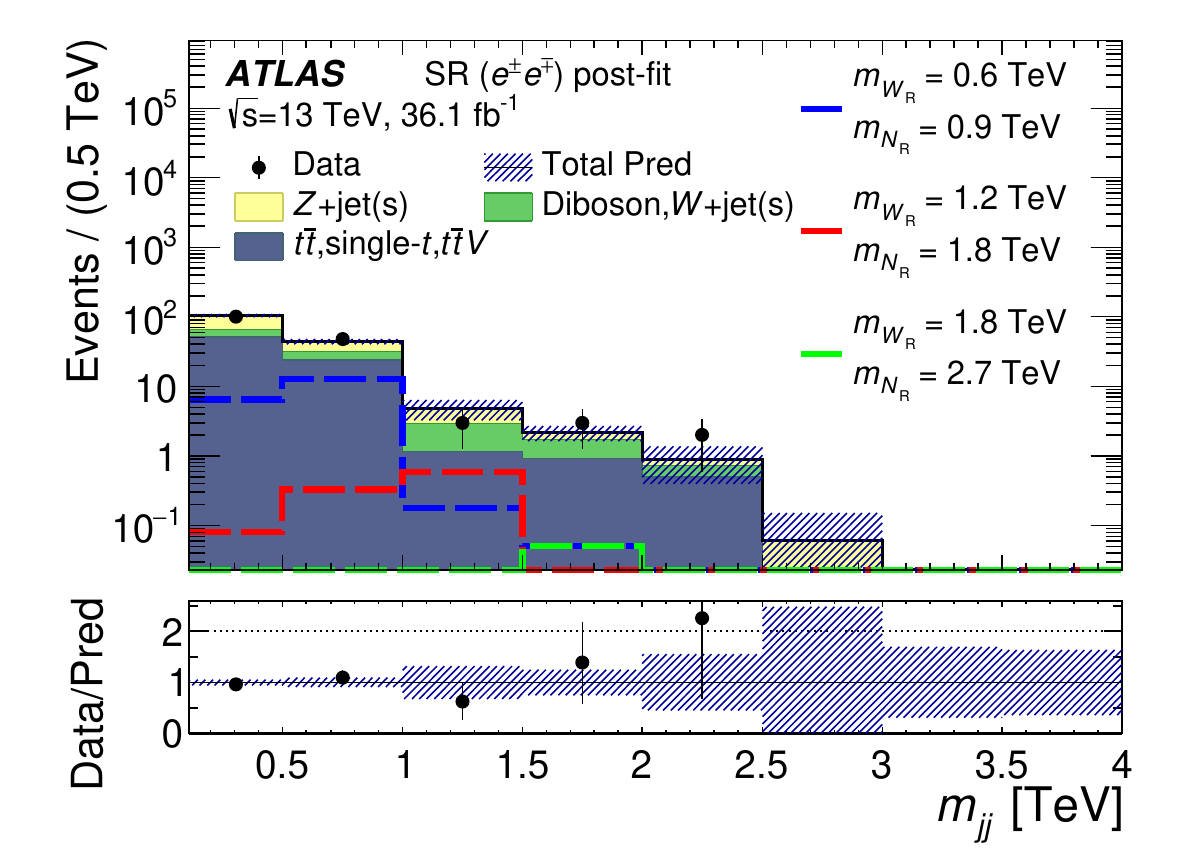}
\label{fig:histo_SR_eeos_mjj_excl_post}
}
\\
\subfloat[]{
\includegraphics[width=0.48\textwidth]{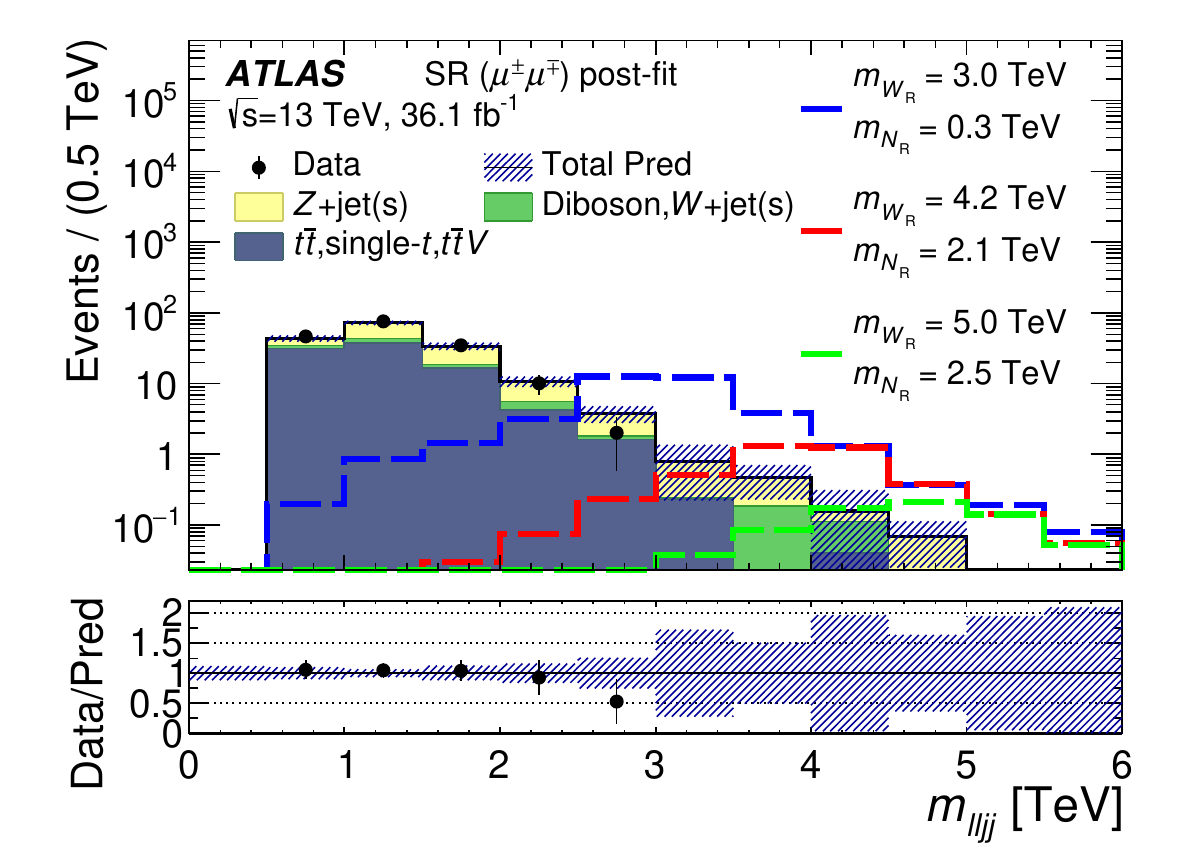}
\label{fig:histo_SR_mmos_excl_post}
}
\subfloat[]{
\includegraphics[width= 0.48\textwidth]{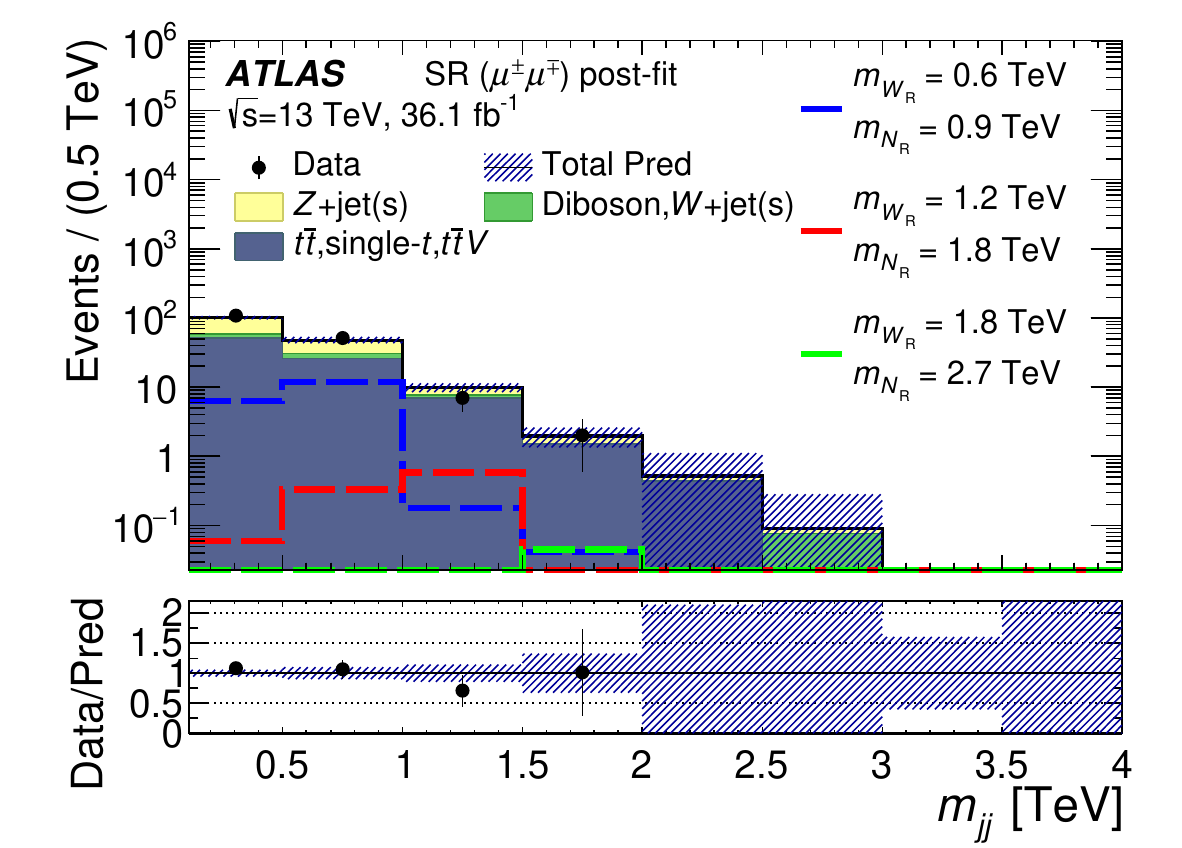}
\label{fig:histo_SR_mmos_mjj_excl_post}
}
\\
\subfloat[]{
\includegraphics[width=0.48\textwidth]{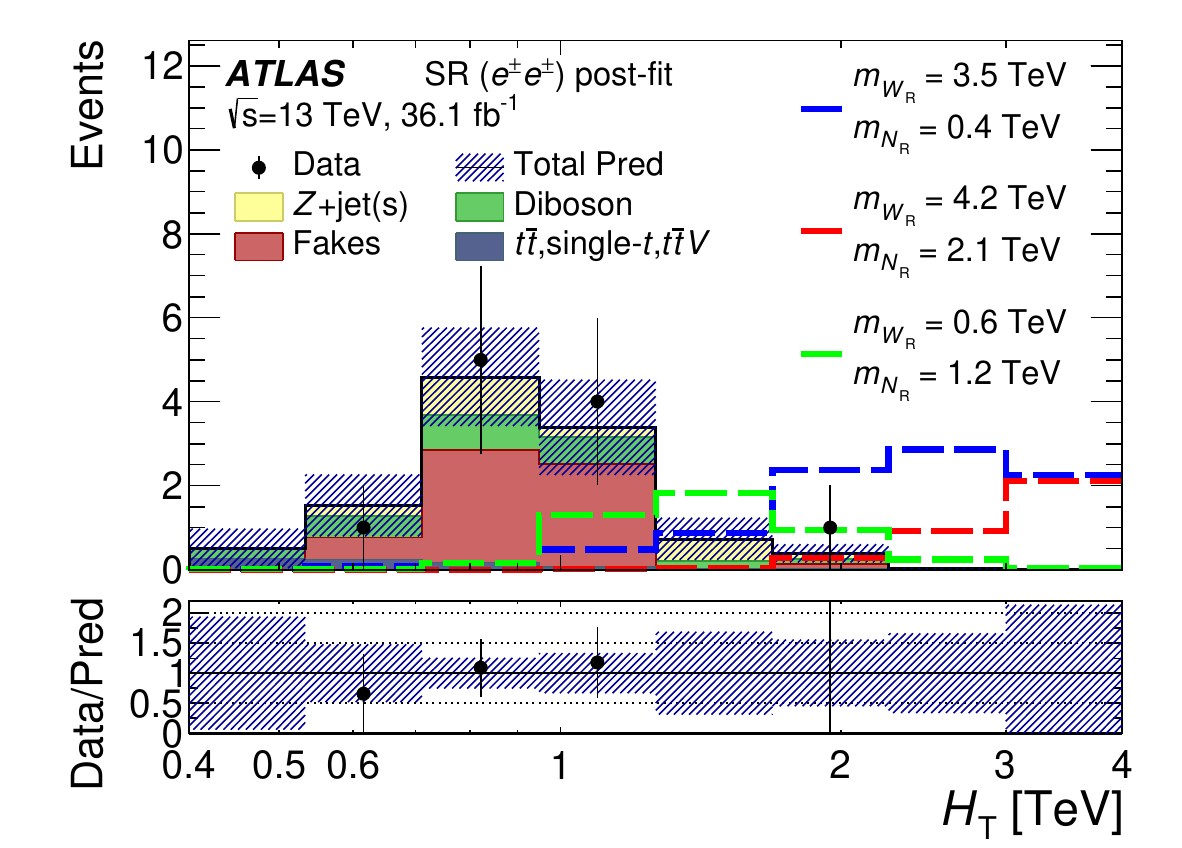}
\label{fig:histo_SSZSRee_excl_post}
}
\subfloat[]{
\includegraphics[width= 0.48\textwidth]{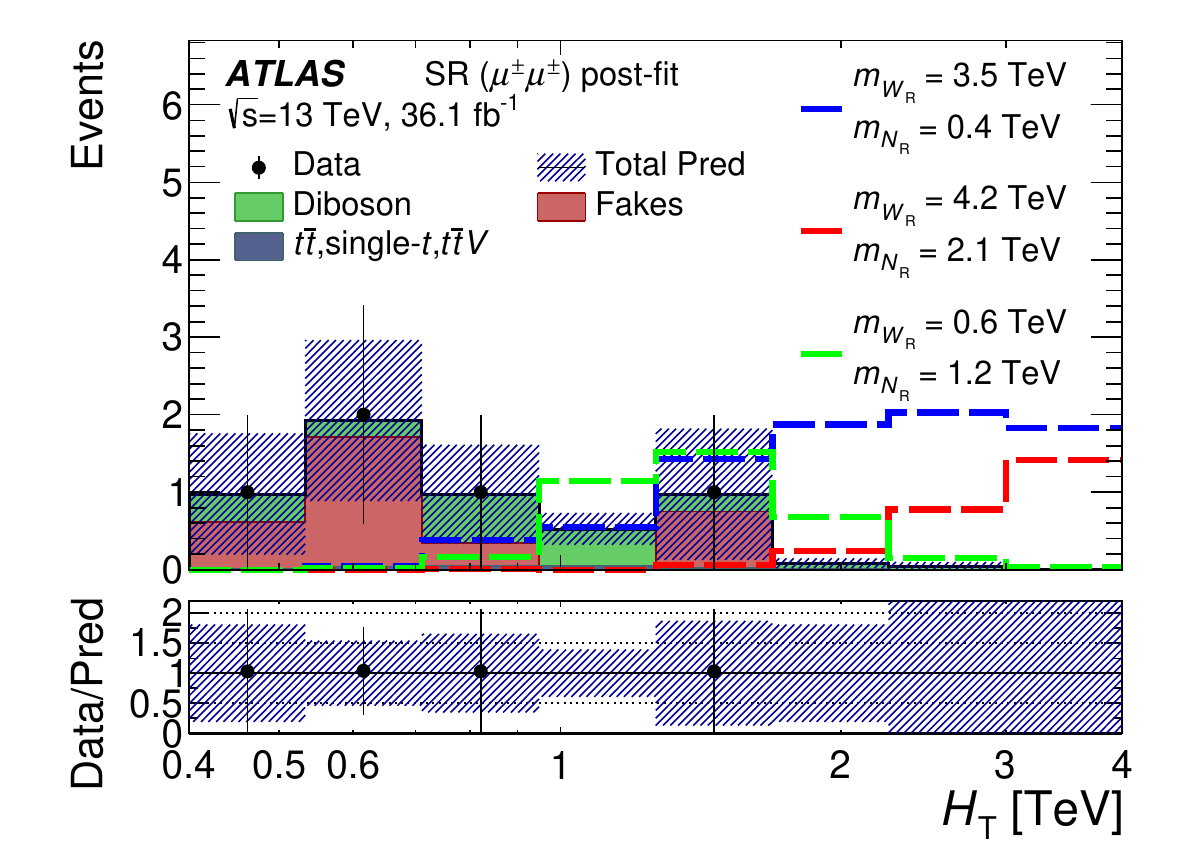}
\label{fig:histo_SSZSRmm_excl_post}
}
\caption{Distributions for data and background predictions after the CR+SR
  fit for discriminant variables in various signal regions and channels:
\protect\subref{fig:histo_SR_eeos_excl_post} \mlljj in \SROSEE,
\protect\subref{fig:histo_SR_eeos_mjj_excl_post} \mjj in \SROSEE,
\protect\subref{fig:histo_SR_mmos_excl_post} \mlljj in \SROSMM,
\protect\subref{fig:histo_SR_mmos_mjj_excl_post} \mjj in \SROSMM,
\protect\subref{fig:histo_SSZSRee_excl_post} \HT in \SRSSEE, and
\protect\subref{fig:histo_SSZSRmm_excl_post} \HT in \SRSSMM.
The hatched bands include all post-fit systematic uncertainties. The correlation between sources of systematic uncertainties are taken into account. A few simulated signal distributions, normalised
to the predicted cross-section, are overlaid on top of the background
prediction in the plots. The last bin in each histogram contains the overflow.}
\label{fig:signalRegions_excl}
\end{figure}

To check the validity of the background prediction in the validation regions
and to estimate the SM background in the signal regions, a background-only fit is performed where the nuisance parameters are constrained only in the CRs and are extrapolated to VRs after the fit.
Figures~\ref{fig:controlRegions_bkg}~and~\ref{fig:validationRegions_bkg} show the distributions in the CRs and VRs after this background-only fit (the binning used in Figures corresponds to the one used in the fit).
When performing the CR+SR fit to obtain the upper limit on the signal strength, 
the nuisance parameters are constrained both in CRs and SRs. For example,
the \SC diboson normalisation is constrained using both \CRSSEE and
\CRSSMM. This diboson normalisation factor is validated in the \VRSSMM, as
shown in Figure~\ref{fig:histo_SSZVRmm_post}, where the expected
background and observed number of events agree within the uncertainties.
In the fit, the Majorana and the Dirac case
are distinguished by using both the \OC and \SC regions or only the \OC regions, respectively.
The SR distributions after the CR+SR fit are shown in
Figure~\ref{fig:signalRegions_excl}.

No significant deviation from the SM predictions is observed in any of the signal regions.
The most significant local excess is observed in the \mlljj spectrum of the \EEOC channel, 
with a $\sim$2$\sigma$ local significance ($\sim$1$\sigma$ global significance) between 3.5 and 4 
\TeV. 
After the fit, the compatibility between the data and the expected
background is assessed. 
The global $p$-values for the background-only hypothesis are 0.67 and 0.62 for the \EESC 
and \MMSC channels respectively and 0.16 and 0.59 for the \EEOC and \MMOC channels.
Upper limits at $95\%$~CL are calculated for the KS process cross-section using the $\text{CL}_{\text{s}}$ 
method~\cite{read2002presentation} and the profile likelihood-ratio~\cite{Cowan:2010js} as the test statistic.
Exclusion limits are calculated in separate fits for the \EE and \MM
channels since there is no theoretical requirement that
\Ne and \Nm have the same mass.

\begin{figure}[htbp]
\centering
\subfloat[]{
\includegraphics[width=0.48\textwidth]{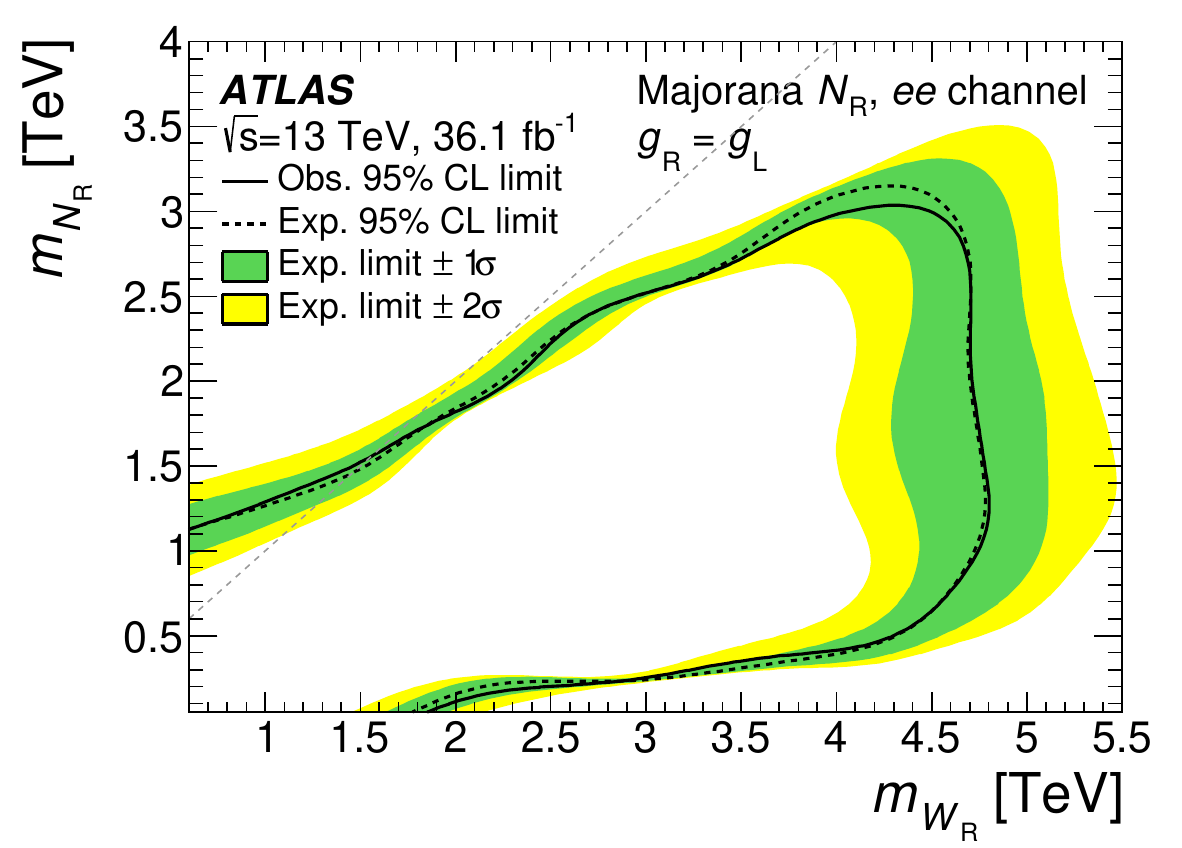}
\label{fig:exclusion_ee}
}
\subfloat[]{
\includegraphics[width= 0.48\textwidth]{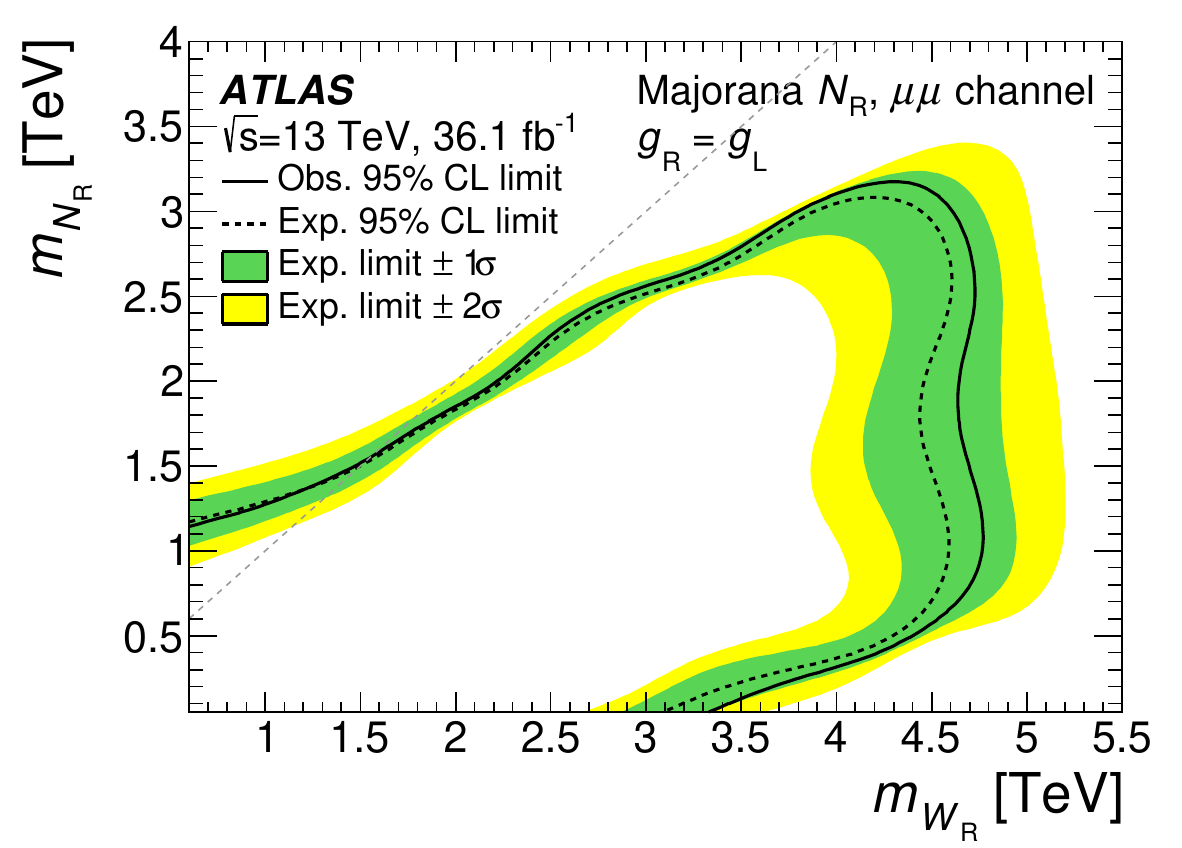}
\label{fig:exclusion_mm}
}
\\
\subfloat[]{
\includegraphics[width=0.48\textwidth]{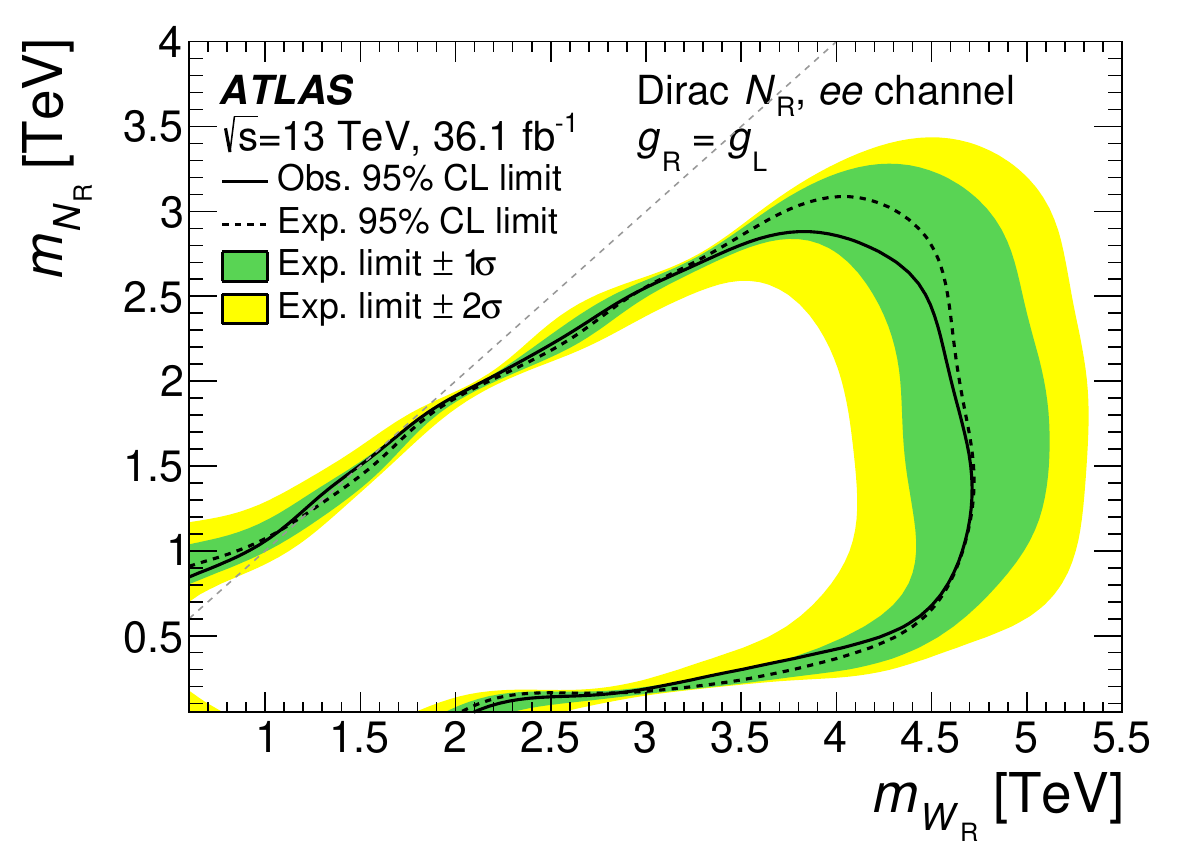}
\label{fig:exclusion_ee_mu_sig2}
}
\subfloat[]{
\includegraphics[width=0.48\textwidth]{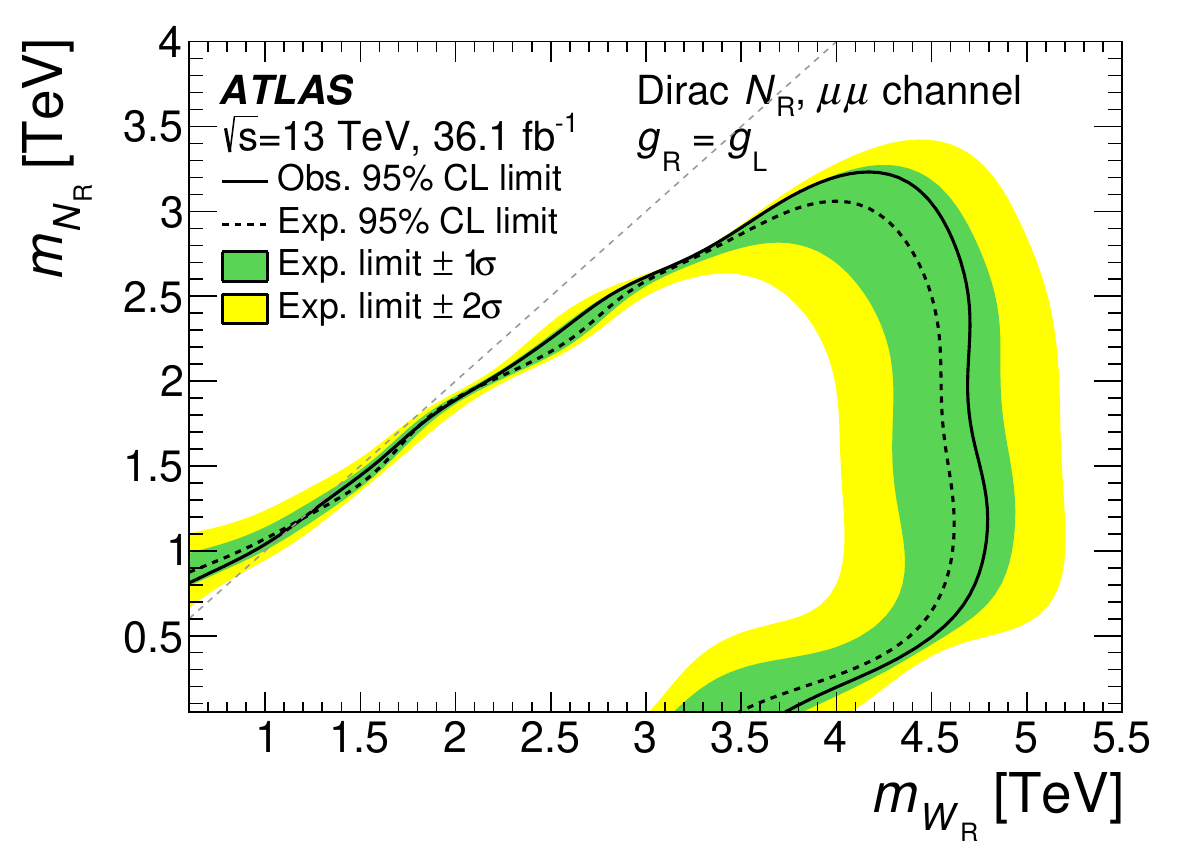}
\label{fig:exclusion_mm_mu_sig2}
}
\caption{
Observed (continuous line) and expected (dashed line) $95\%$ CL exclusion contours
in the \mWR--\mNR plane, along with the one- and two-standard-deviation uncertainty 
bands around the expected exclusion contour in the
\protect\subref{fig:exclusion_ee} \EE and 
\protect\subref{fig:exclusion_mm} \MM channels for Majorana \NR neutrinos,
\protect\subref{fig:exclusion_ee_mu_sig2} \EE and
\protect\subref{fig:exclusion_mm_mu_sig2} \MM channels for Dirac \NR neutrinos.
The dashed gray line indicates the region of the plane where $\mWR = \mNR$.
The left- and right-handed weak gauge couplings are assumed to be the same, as indicated by $g_{\text{L}} = g_{\text{R}}$.
}
\label{fig:Exclusion}
\end{figure}

\begin{figure}[htbp]
\centering
\subfloat[]{
\includegraphics[width=0.48\textwidth]{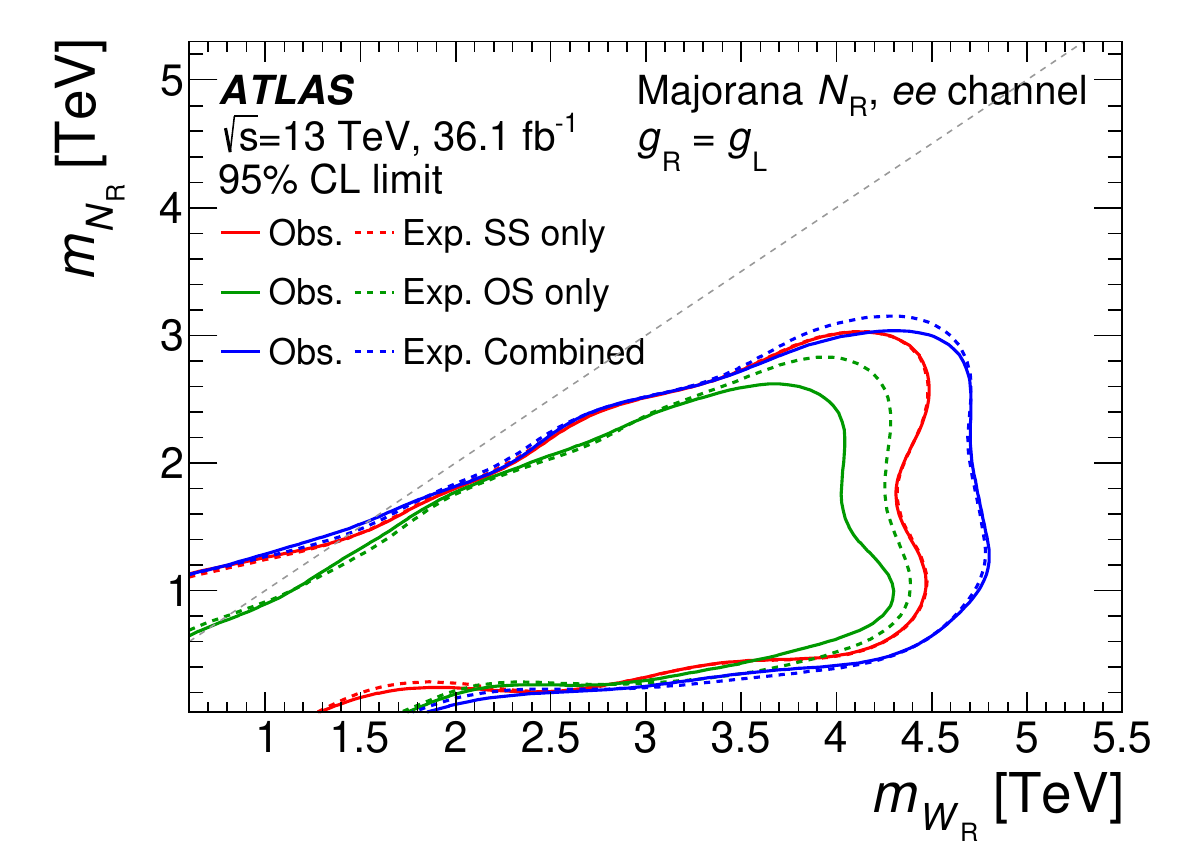}
\label{fig:exclusion_ee_OS}
}
\subfloat[]{
\includegraphics[width= 0.48\textwidth]{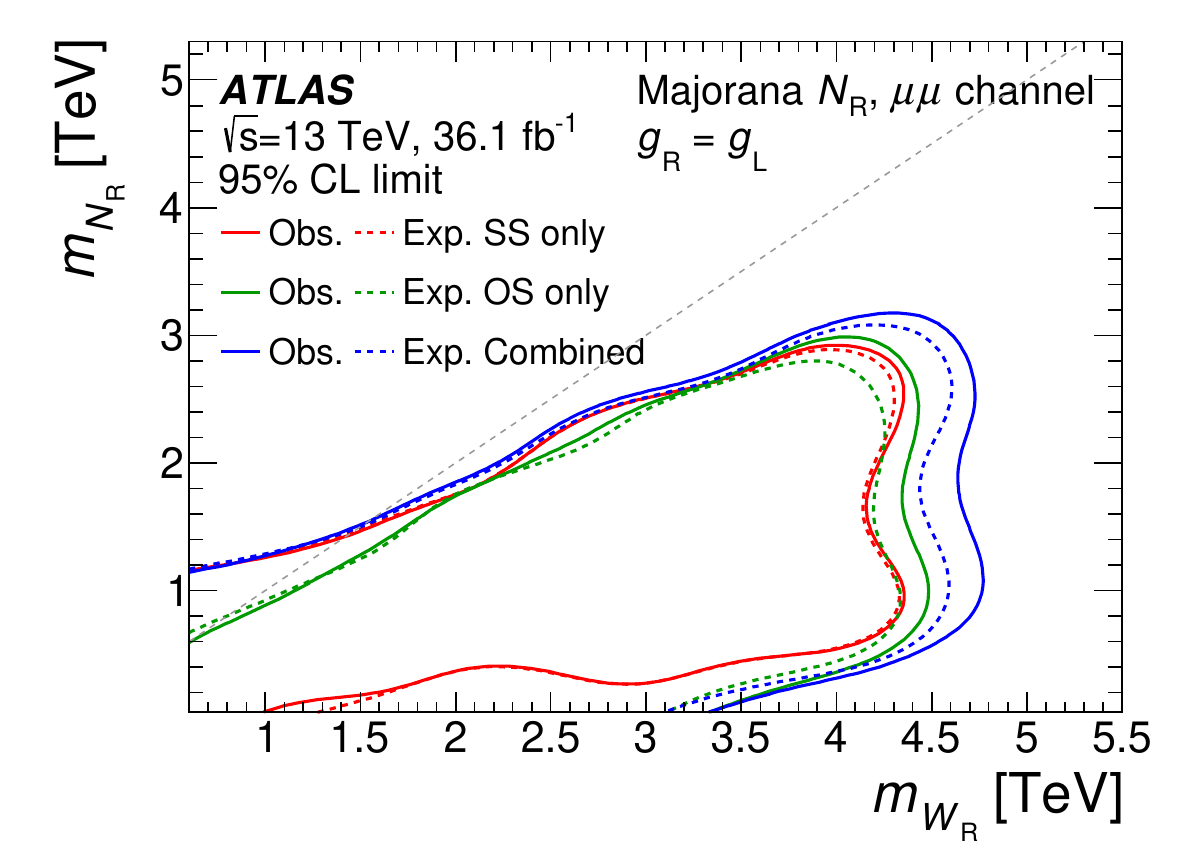}
\label{fig:exclusion_mm_OS}
}
\caption{
Observed (continuous line) and expected (dashed line) $95\%$ CL exclusion
contours in the \mWR--\mNR plane for Majorana \NR neutrinos calculated in
the \OC and \SC analyses, and their combination, in the
\protect\subref{fig:exclusion_ee_OS} \EE and
\protect\subref{fig:exclusion_mm_OS} \MM channels.
The dashed gray line indicates the region of the plane where $\mWR = \mNR$.
The left- and right-handed weak gauge couplings are assumed to be the same ($g_\mathrm{L} = g_\mathrm{R}$).}
\label{fig:Exclusion_sep}
\end{figure}

When setting exclusion limits for the two models
the Dirac case is assumed to give twice as many \OC events as the Majorana case.
Results are shown in Figure~\ref{fig:Exclusion} for both Majorana and Dirac
\NR neutrino cases.
In both cases, \WR bosons with masses up to 4.7~\TeV{} are excluded at
$95\%$~CL, for \mNR in the \SI{0.5}-\SI{3.0}{\TeV} region. 
The two jets in the \WR boson decay chain, for $m_{W_{\mathrm{R}\xspace}}{\sim}$\SI{3}-\SI{3.5}{\TeV} and \mNR neutrino masses 
 below \SI{0.5}{\TeV}, tend to overlap, resulting in a rapidly changing sensitivity 
and narrow 1$\sigma$ and 2$\sigma$ uncertainty bands in the exclusion limit plot, as visible in Figures~\ref{fig:exclusion_ee} and~\ref{fig:exclusion_ee_mu_sig2}.
Figure~\ref{fig:Exclusion_sep} shows the exclusion limits separately for
the \OC and \SC analyses in the Majorana \NR neutrino case.
The two analyses generally exhibit a similar sensitivity across the
two-dimensional mass plane. Under the inverted hierarchy hypothesis ($\mNR
> \mWR$) the \SC analysis is more sensitive due to the lower expected
background (Figures \ref{fig:histo_SR_eeos_mjj_excl_post},
\ref{fig:histo_SR_mmos_mjj_excl_post}, \ref{fig:histo_SSZSRee_excl_post} and
\ref{fig:histo_SSZSRmm_excl_post}).  

In the electron channel (Fig.~\ref{fig:exclusion_ee_OS}), the
\OC observed limit around $m_{W_{\mathrm{R}\xspace}}{\sim}$\SI{4.2}{\TeV} and
$m_{N_{\mathrm{R}\xspace}}{\sim}$\SI{3}{\TeV} is weaker than expected due to three events observed at 
$m_{\ell\ell jj}{\sim}$\SI{3.8}{\TeV}, compared to an expected background of $1.1\pm0.4$ events
for \mlljj $>$\SI{3.5}{\TeV} (Fig.~\ref{fig:histo_SR_eeos_excl_post}). The
opposite effect is observed in the muon channel
(Fig.~\ref{fig:exclusion_mm_OS}), where no events are observed in the
$\mlljj$ $>$\SI{3}{\TeV} region (Fig.~\ref{fig:histo_SR_mmos_excl_post}). 

\FloatBarrier

\FloatBarrier

\FloatBarrier

\section{Conclusion}
\label{sec:conclusion}
A search for right-handed $W$ bosons and heavy right-handed Majorana or
Dirac neutrinos is presented using a final state
containing a pair of charged leptons, electrons or muons, and two jets (\lljj), with $\ell = e,
\mu$, in a 36.1~\ifb{} sample of $pp$ collisions recorded by the ATLAS detector at
$\sqrt{s} = 13~\TeV{}$ at LHC. No evidence of \WR bosons or Majorana or Dirac heavy
neutrinos, \NR, is found assuming the KS production and lower limits are set on the \mWR and \mNR masses, assuming equality of left- and right-handed weak gauge couplings ($g_{\text{L}} = g_{\text{R}}$). 
The excluded region for the Majorana \NR neutrinos extends to about $\mWR=\SI{4.7}{\TeV}$, for $\mNR=\SI{1.2}{\TeV}$ in the electron channel and for $\mNR=\SI{1}{\TeV}$ in the muon channel. 
The \mNR limits reach about \SI{2.9}{\TeV} in the electron channel and \SI{3.1}{\TeV} in the muon channel (for $\mWR=\SI{4.3}{\TeV}$).
For Dirac \NR neutrinos, limits reach about $\mWR=\SI{4.7}{\TeV}$, for $\mNR=\SI{1}{\TeV}$ in the electron channel and for $\mNR=\SI{1.2}{\TeV}$ in the muon channel.
Limits on \mNR up to about \SI{2.8}{\TeV} (for $\mWR=\SI{3.7}{\TeV}$) in the electron channel and up to \SI{3.2}{\TeV} (for $\mWR=\SI{4.1}{\TeV}$) in the muon channel are set.
In the low-mass regime ($\mWR < \SI{1.5}{\TeV}$), under the hierarchy hypothesis \mNR $>$ \mWR, \NR masses up to ~1.5~\TeV{} are excluded at 95\% CL.
These results improve upon previous ATLAS searches~\cite{EXOT-2012-24} and
extend the exclusion limits on \mWR by 1--\SI{2}{\TeV}. Additionally, the
scenario in which the \NR neutrino is heavier than the \WR boson is
explored for the first time.

\FloatBarrier
\clearpage
\section*{Acknowledgements}

We thank CERN for the very successful operation of the LHC, as well as the
support staff from our institutions without whom ATLAS could not be
operated efficiently.

We acknowledge the support of ANPCyT, Argentina; YerPhI, Armenia; ARC, Australia; BMWFW and FWF, Austria; ANAS, Azerbaijan; SSTC, Belarus; CNPq and FAPESP, Brazil; NSERC, NRC and CFI, Canada; CERN; CONICYT, Chile; CAS, MOST and NSFC, China; COLCIENCIAS, Colombia; MSMT CR, MPO CR and VSC CR, Czech Republic; DNRF and DNSRC, Denmark; IN2P3-CNRS, CEA-DRF/IRFU, France; SRNSFG, Georgia; BMBF, HGF, and MPG, Germany; GSRT, Greece; RGC, Hong Kong SAR, China; ISF and Benoziyo Center, Israel; INFN, Italy; MEXT and JSPS, Japan; CNRST, Morocco; NWO, Netherlands; RCN, Norway; MNiSW and NCN, Poland; FCT, Portugal; MNE/IFA, Romania; MES of Russia and NRC KI, Russian Federation; JINR; MESTD, Serbia; MSSR, Slovakia; ARRS and MIZ\v{S}, Slovenia; DST/NRF, South Africa; MINECO, Spain; SRC and Wallenberg Foundation, Sweden; SERI, SNSF and Cantons of Bern and Geneva, Switzerland; MOST, Taiwan; TAEK, Turkey; STFC, United Kingdom; DOE and NSF, United States of America. In addition, individual groups and members have received support from BCKDF, CANARIE, CRC and Compute Canada, Canada; COST, ERC, ERDF, Horizon 2020, and Marie Sk{\l}odowska-Curie Actions, European Union; Investissements d' Avenir Labex and Idex, ANR, France; DFG and AvH Foundation, Germany; Herakleitos, Thales and Aristeia programmes co-financed by EU-ESF and the Greek NSRF, Greece; BSF-NSF and GIF, Israel; CERCA Programme Generalitat de Catalunya, Spain; The Royal Society and Leverhulme Trust, United Kingdom. 

The crucial computing support from all WLCG partners is acknowledged gratefully, in particular from CERN, the ATLAS Tier-1 facilities at TRIUMF (Canada), NDGF (Denmark, Norway, Sweden), CC-IN2P3 (France), KIT/GridKA (Germany), INFN-CNAF (Italy), NL-T1 (Netherlands), PIC (Spain), ASGC (Taiwan), RAL (UK) and BNL (USA), the Tier-2 facilities worldwide and large non-WLCG resource providers. Major contributors of computing resources are listed in Ref.~\cite{ATL-GEN-PUB-2016-002}.



\printbibliography


\clearpage 
 
\begin{flushleft}
{\Large The ATLAS Collaboration}

\bigskip

M.~Aaboud$^\textrm{\scriptsize 34d}$,    
G.~Aad$^\textrm{\scriptsize 99}$,    
B.~Abbott$^\textrm{\scriptsize 125}$,    
O.~Abdinov$^\textrm{\scriptsize 13,*}$,    
B.~Abeloos$^\textrm{\scriptsize 129}$,    
D.K.~Abhayasinghe$^\textrm{\scriptsize 91}$,    
S.H.~Abidi$^\textrm{\scriptsize 164}$,    
O.S.~AbouZeid$^\textrm{\scriptsize 39}$,    
N.L.~Abraham$^\textrm{\scriptsize 153}$,    
H.~Abramowicz$^\textrm{\scriptsize 158}$
% ...
\bigskip
\\

% ...
$^{13}$Institute of Physics, Azerbaijan Academy of Sciences, Baku; Azerbaijan.\\
% ...
$^{34}$$^{(a)}$Facult\'e des Sciences Ain Chock, R\'eseau Universitaire de Physique des Hautes Energies - Universit\'e Hassan II, Casablanca;$^{(b)}$Centre National de l'Energie des Sciences Techniques Nucleaires (CNESTEN), Rabat;$^{(c)}$Facult\'e des Sciences Semlalia, Universit\'e Cadi Ayyad, LPHEA-Marrakech;$^{(d)}$Facult\'e des Sciences, Universit\'e Mohamed Premier and LPTPM, Oujda;$^{(e)}$Facult\'e des sciences, Universit\'e Mohammed V, Rabat; Morocco.\\
% ...
$^{39}$Niels Bohr Institute, University of Copenhagen, Copenhagen; Denmark.\\
% ...
$^{91}$Department of Physics, Royal Holloway University of London, Egham; United Kingdom.\\
% ...
$^{99}$CPPM, Aix-Marseille Universit\'e, CNRS/IN2P3, Marseille; France.\\
% ...
$^{125}$Homer L. Dodge Department of Physics and Astronomy, University of Oklahoma, Norman OK; United States of America.\\
% ...
$^{129}$LAL, Universit\'e Paris-Sud, CNRS/IN2P3, Universit\'e Paris-Saclay, Orsay; France.\\
% ...
$^{153}$Department of Physics and Astronomy, University of Sussex, Brighton; United Kingdom.\\
% ...
$^{158}$Raymond and Beverly Sackler School of Physics and Astronomy, Tel Aviv University, Tel Aviv; Israel.\\
% ...
$^{164}$Department of Physics, University of Toronto, Toronto ON; Canada.\\
% ...
$^{a}$ Also at Borough of Manhattan Community College, City University of New York, NY; United States of America.\\
$^{b}$ Also at California State University, East Bay; United States of America.\\
$^{c}$ Also at Centre for High Performance Computing, CSIR Campus, Rosebank, Cape Town; South Africa.\\
$^{d}$ Also at CERN, Geneva; Switzerland.\\
$^{e}$ Also at CPPM, Aix-Marseille Universit\'e, CNRS/IN2P3, Marseille; France.\\
% ...
$^{*}$ Deceased

\end{flushleft}


\end{document}